%% file: main.tex
\documentclass[compsoc,conference,a4paper]{IEEEtran}
\usepackage{mypreamble}

\begin{document}

\title{\shortacronym: Dynamically Configured LLM-based Honeypot for Industrial Protocol and Physical Process Emulation}

\author{
\IEEEauthorblockN{
Christoforos Vasilatos,
Dunia J. Mahboobeh,
Hithem Lamri,
Manaar Alam, and
Michail Maniatakos
}
\IEEEauthorblockA{
Center for Cyber Security, New York University Abu Dhabi, UAE\\
\{cv43, dam10098, hl5743, alam.manaar, mihalis.maniatakos\}@nyu.edu
}
}

\maketitle
\begin{abstract}
\input{sections/abstract}
\end{abstract}

\section{Introduction}
\input{sections/01-introduction}

\section{Background}
\input{sections/02-background}

\section{Preliminaries}\label{section:preliminaries}
\input{sections/03-preliminaries}

\section{\shortacronym~Overview}\label{section:overview}
\input{sections/04-overview}

\section{\shortacronym~ICS Network Protocol Emulation}\label{section:protocol_emulation}
\input{sections/05-protocol_emulation}

\section{\shortacronym~Physical Process Emulation}\label{section:process_emulation}
\input{sections/06-process_emulation}

\section{Case Study: Desalination Plant Testbed}\label{section:process_testbed}
\input{sections/07-case_study}

\section{Comparison to Related Work}\label{section:related_work}
\input{sections/08-related_work}

\section{Honeypot Development and Analysis}\label{section:honeypot_setup}
\input{sections/09-honeypot_setup}

\section{Discussion and Limitations}
\input{sections/10-discussion}

\section{Conclusion}
\input{sections/11-conclusion}

\section*{Acknowledgments} This work has been supported by the NYUAD Center for Cyber Security under RRC Grant No. G1104.

\section*{Availability} The source code, models, and datasets are open source at \url{https://github.com/momalab/LLMPot}.

\bibliographystyle{IEEEtranS}
\bibliography{references}

\appendix
\input{sections/appendix}

\end{document}

%% file: sections/abstract.tex
Industrial Control Systems (ICS) are extensively used in critical infrastructures ensuring efficient, reliable, and continuous operations. However, their increasing connectivity and addition of advanced features make them vulnerable to cyber threats, potentially leading to severe disruptions in essential services. In this context, honeypots play a vital role by acting as decoy targets within ICS networks, or on the Internet, helping to detect, log, analyze, and develop mitigations for ICS-specific cyber threats. Deploying ICS honeypots, however, is challenging due to the necessity of accurately replicating industrial protocols and device characteristics, a crucial requirement for effectively mimicking the unique operational behavior of different industrial systems. Additionally, the difficulty is increased by the substantial manual effort involved in replicating the PLC's control logic. This is necessary to capture attacker traffic that seeks to interfere with critical infrastructure operations. In this paper, we propose \shortacronym, a novel approach for designing honeypots in ICS networks harnessing the potency of Large Language Models (LLMs). \shortacronym~aims to provide a dynamic framework that can be used to optimize the creation of realistic honeypots with vendor-agnostic configurations and for various control logic, aiming to eliminate the manual effort and specialized knowledge traditionally required by existing strategies. We conducted extensive experiments focusing on a wide array of parameters, demonstrating that \shortacronym~can effectively create honeypot devices implementing different industrial protocols, PLC configurations, and diverse control logic.

%% file: sections/01-introduction.tex
Contemporary industrial frameworks incorporate vital infrastructures, such as Programmable Logic Controllers (PLCs), which serve as integral components of Industrial Control Systems (ICS) deployed across various sectors including power generation and distribution, water treatment plants, and manufacturing facilities, among others. Because of the essential role these infrastructures play, ICS have become a frequent target for cyber attacks \cite{mclaughlin2016cybersecurity}, while at the same time researchers are developing tools to uncover vulnerabilities~\cite{tychalas2021icsfuzz, bytes2023fieldfuzz, icsquartz2025ndss}. With the heightened interconnectivity of ICS devices and the expanded attack surface resulting from the numerous capabilities provided by modern ICS, attackers have numerous potential entry points into these systems.

A honeypot serves as a security mechanism specifically designed in a controlled and isolated environment to improve the cyber security posture of ICS devices~\cite{piggin2016active, campbell2015survey, moore2016detecting, weiler2002honeypots}. The objective of honeypots is to mimic the functionalities of real ICS devices so convincingly that they attract and redirect malicious activities away from legitimate devices. This allows security professionals to monitor and analyze intercepted network traffic, providing insight into attacker methods and improving the security of critical infrastructures. Therefore, the primary challenge lies in developing a honeypot that closely mimics the behavior of ICS devices with such precision that it becomes indistinguishable from the actual devices, ensuring its effectiveness as a security decoy. 

\subsubsection*{Limitations and Challenges} Existing ICS honeypot frameworks primarily focus on emulating industrial protocols of specific PLC devices or given PLC setups. That is, \emph{predefined} analog and digital inputs are available and/or a predefined control logic that the PLC is capturing. In order to develop such a honeypot, extensive domain knowledge and implementation capabilities are required; therefore, related works base their honeypots on existing frameworks for protocol emulation~\cite{jicha2016scada,antonioli2015minicps}. Besides the industrial protocol emulation, ICS would typically control some physical process, therefore realistic honeypots should also support some form of emulation of an underlying physical process. Related work either does not support physical process emulation~\cite{cao2018dipot,xiao2018s7commtrace,lopez2020honeyplc,siniosoglou2020neuralpot}, or their capabilities tend to be \emph{restricted to very specific scenarios} (e.g., water treatment processes~\cite {antonioli2016towards}) and \emph{lack the flexibility to modify or extend the emulated control logic}~\cite{litchfield2016rethinking,conti2022icspot}. Current emulation strategies, which \emph{rely on static scripting or fully-fledged applications}~\cite{lopez2020honeyplc}, often fail to cover all possible scenarios and do not give the user the ability to alter the behavior of the honeypot. Therefore, the main challenge of efficient honeypot development is the tedious manual work needed to emulate the hardware, the industrial protocol that depicts the unique PLC configuration, and the control logic. Especially the protocol emulation needs deep understanding of protocol manuals and details that are hard to be depicted accurately and exhaustively in a rule-based program.
    
\subsubsection*{Why Large Language Models (LLMs)} While LLMs have primarily been applied to natural language and code generation, their potential for generating network traffic (bytes) remains largely unexplored. LLMs excel at generating sequences of responses to queries. In this context, queries and responses correspond to network packets. With proper training, LLMs can generate precise, context-aware network responses, making them ideal core components for ICS honeypots. In other words, LLMs have the potential to significantly reduce the manual effort required to emulate hardware, protocols, and industrial/physical processes in honeypot development. Our framework establishes a specialized interaction with PLCs to generate customized datasets optimized for training LLMs as network emulators. Traditional methods often require implementing detailed physical process simulations using complex tools such as Simulink/MATLAB, Hardware-in-the-Loop (HIL) simulations, and extensive client-side implementations for testing.~\shortacronym~takes a different approach: instead of relying on pre-existing physical process models, it collects traffic by probing PLCs with targeted requests using only the client-side implementation, decoupling from server complexity. This probing can be performed on both simulated and real-world physical processes. This method offers significant advantages, including adaptability to diverse physical process emulation and freeing engineers and researchers from the time-consuming task of manually developing process simulations.

Therefore, in this paper, we introduce~\shortacronym, an LLM-based approach for protocol and physical process emulation, leveraging pre-trained LLMs. To the best of our knowledge, this is the first work where pre-trained LLMs are used to emulate ICS protocols, physical processes, and control logic installed in PLCs. The use of pre-trained machine learning models offers two advantages. First, it significantly reduces resource consumption and eliminates the need for costly and time-intensive training from scratch. Second, these pre-trained LLMs can be efficiently fine-tuned with a relatively small and domain-specific dataset, enhancing their ability to generate specialized and accurate responses. 

\begin{figure}[!t]
\centering
\includegraphics[width=\linewidth]{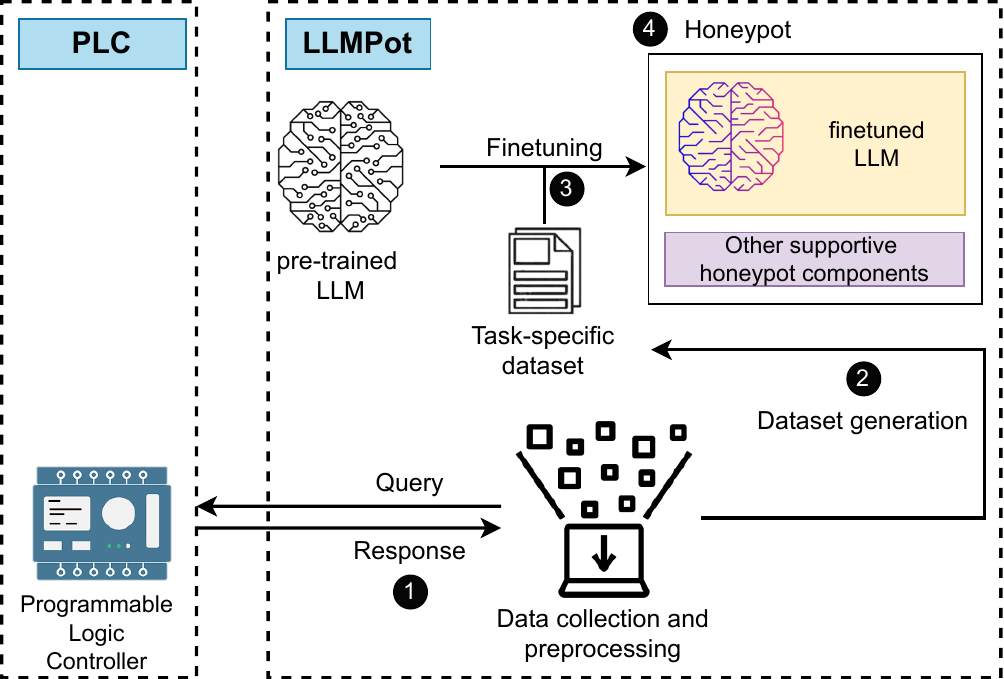}
\caption{\shortacronym~high-level diagram. 1. A client that automatically probes the PLC and captures responses. 2. Captured traffic forms a training dataset. 3. Fine-tuning of the LLM using the dataset. 4. Generated LLM-based honeypot with supportive components.}
\label{fig:high-level}
\end{figure}

In conclusion, this effort aims to enable users to customize the analog/digital configuration or the control logic running on a PLC on demand and thus create a more realistic and adaptive honeypot that meets the user's requirements. As Figure~\ref{fig:high-level} shows, our methodology enables \textbf{``cloning''} of an industrial protocol given a specific PLC configuration on a specific port (example, ``Modbus:502'') and instantiates it as an  \shortacronym~instance. After \shortacronym~has been successfully trained, it can be deployed as a single or multiple instances with different public IPs to capture diverse attacker traffic. The PLC device is used only during the offline phase to build the honeypot and is not needed during the honeypot deployment.

Our contributions can be summarized as follows:
\begin{itemize}[leftmargin=*]
    \item A novel dynamic framework for generating a customized dataset that \textbf{captures the PLC device-specific setup including the analog and digital configuration}. The approach is flexible and can support a diverse set of industrial protocols.
    \item A platform for fine-tuning LLMs using the previously generated dataset, resulting in a cloned PLC device that accurately emulates both the ICS network protocol and control logic. This fine-tuned model can then serve as the emulator module within an ICS honeypot setup.
    \item Three new metrics for evaluating the performance of LLMs as ICS emulators: \textit{Byte-to-byte Comparison Accuracy} (BCA), \textit{Response Validity Accuracy} (RVA), and \textit{Response Validity Accuracy - Epsilon} (\rvae).
    \item The source code, models, and datasets are open source at \url{https://github.com/momalab/LLMPot}.
\end{itemize}

%% file: sections/02-background.tex
A key component in ICS networks is the PLC, a compact industrial computer programmed to perform logical operations based on inputs from sensors, relays, switches, and others to control actuators, valves, and more. Prominent PLC vendors include Allen-Bradley, Siemens, WAGO, and ABB. ICS communication protocols like Modbus~\cite{modbus}, S7comm~\cite{s7comm}, EtherCAT~\cite{ethercat}, and Profinet~\cite{profinet} have been developed to enable reliable and efficient data exchange within these PLCs to monitor physical processes across various sectors.

Table~\ref{table:modbus_s7-frame} depicts the packet frame formats for two different industrial communication protocols: Modbus and S7comm. The standard Modbus protocol is wrapped in a TCP/IP layer, enabling communications over Ethernet networks. Modbus query~\blackcircle{a} and response~\blackcircle{b} packet frames contain a Modbus Application Protocol (MBAP) header and a Protocol Data Unit (PDU). On the other hand, the S7comm query~\blackcircle{c} and response~\blackcircle{d} packet frames generally include a header, a parameter, and a data field. Headers in both protocols are crucial for correlating queries with responses by matching several identifiers (e.g., Transaction ID, Protocol ID, etc.) and verifying frame length. A key element in both protocols is the function code (FC) in the PDU or parameter section, which dictates the command or operation a client requests to the server. The data fields, which depend on the FC, detail the variables involved and their count. In both protocols, the data is represented as a series of digital and analog input/output; therefore, both protocols support functions for reading and writing to these data types.

\begin{table}[!t]
    \centering
    \caption{Modbus and S7 Communication Frames}
    \label{table:modbus_s7-frame}

    \resizebox{\columnwidth}{!}{%
    \begin{tabular}{p{0.3cm}|c|c|c|c|c|c|}  
        \cline{2-7}
        & \multicolumn{6}{c|}{\thead{\textbf{Modbus Communication}}} \\
        \cline{2-7}
        & \multicolumn{4}{c|}{\thead{\textbf{Modbus Application Protocol (MBAP) Header}}} & \multicolumn{2}{c|}{\thead{\textbf{Protocol Data Unit (PDU)}}} \\
        \cline{2-7}
        & Transaction ID & Protocol ID & Length & Unit ID & Function Code (FC) & Data \\
        \cline{2-7}
        \blackcircle{a} & 00 01 & 00 00 & 00 06 & 00 & 03 & 00 01 00 01 \\
        \cline{2-7}
        \blackcircle{b} & 00 01 & 00 00 & 00 05 & 00 & 03 & 02 00 00 \\
        \cline{2-7}
    \end{tabular}
    }
    
    \vspace{0.1cm}

    \resizebox{\columnwidth}{!}{%
    \begin{tabular}{p{0.3cm}|c|c|c|c|c|c|c|c|c|c|}
        \cline{2-10}
        & \multicolumn{9}{c|}{\thead{\textbf{S7 Communication - Query}}} \\
        \cline{2-10}
        & \multicolumn{6}{c|}{\thead{\textbf{Header}}} & \multicolumn{3}{c|}{\thead{\textbf{Parameter}}} \\
        \cline{2-10}
        & \makecell{Protocol\\ID} & ROSCTR & \makecell{Redundancy\\ID} & \makecell{(PDU)\\Reference} & \makecell{Parameter\\length} & \makecell{Data\\length} & \makecell{Function\\Code (FC)} & \makecell{Item\\count} & Items \\ 
        \cline{2-10}
        \blackcircle{c} & 32 & 01 & 00 00 & 01 00 & 00 0e & 00 00 & 04 & 01 & 12 0a .. \\
        \cline{2-10}
    \end{tabular}
    }

    \vspace{0.1cm}

    \resizebox{\columnwidth}{!}{%
    \begin{tabular}{p{0.3cm}|c|c|c|c|c|c|c|c|c|c|c|}
        \cline{2-11}
        & \multicolumn{10}{c|}{\thead{\textbf{S7 Communication - Response}}} \\
        \cline{2-11}
        & \multicolumn{7}{c|}{\thead{\textbf{Header}}} & \multicolumn{3}{c|}{\thead{\textbf{Parameter}}} \\
        \cline{2-11}
        & \makecell{Protocol\\ID} & ROSCTR & \makecell{Redundancy\\ID} & \makecell{(PDU)\\Reference} & \makecell{Parameter\\length} & \makecell{Data\\length} & \makecell{Error\\class/code} & \makecell{Function\\Code (FC)} & \makecell{Item\\count} & Items \\ 
        \cline{2-11}
        \blackcircle{d} & 32 & 03 & 00 00 & 01 00 & 00 02 & 00 04 & 00 00 & 04 & 01 & 05 .. \\
        \cline{2-11}
    \end{tabular}
    }

\end{table}

%% file: sections/03-preliminaries.tex
\subsection{Research Questions}
\label{subsection:research_questions}
Using an LLM to emulate industrial protocols and physical processes involves intricate tasks. For industrial protocols, it is essential to capture and replicate the complex interactions between devices accurately. Likewise, for physical processes, accurately modeling variable changes is important. These efforts require careful data preparation for the model training, which presents various research questions in achieving realism and maintaining accuracy:

\begin{itemize}[leftmargin=*]
    \item \textbf{RQ1: Protocol Emulation Challenge}: How can an LLM emulate the behavior of an ICS protocol with its wide range of functions and values? Can it provide realism by incorporating exception handling, and to what extent? Can the dataset generation be extended to other protocols and dynamically configured to adapt to different PLC configurations and circumstances? [Section~\ref{section:protocol_emulation}] 
    \item \textbf{RQ2: Process Emulation Challenge}: How can an LLM learn to emulate the control logic of industrial physical processes and accurately capture the states of the variables? What is the limit of complexity for control logic that this framework can handle without failing? [Sections~\ref{section:process_emulation} and Section~\ref{section:process_testbed}]
    \item \textbf{RQ3: Honeypot Efficiency}: 
    How effective is the generated honeypot? Does it reflect the PLC device's authentic characteristics upon scanning and disguise itself as a realistic device? Can the honeypot deceive network reconnaissance tools? [Section~\ref{section:honeypot_setup}]
\end{itemize}

\subsection{LLM Evaluation Metrics}\label{subsection:metrics}
A key element of ICS honeypots is their interactivity, which relies on their ability to respond accurately to requests with appropriate data. Since \shortacronym~utilizes machine learning models to emulate the ICS device network communications we define three different metrics in order to assess~\shortacronym's ability to respond appropriately:

\begin{definition}[label={def:accuracy}]{Byte-to-byte Comparison Accuracy (\bca)}
A byte-to-byte comparison of the expected response with the response generated from the fine-tuned LLM. The total accuracy is a percentage of the exact correct responses over the total requests sent.
\end{definition}

The \bca~metric measures whether the LLM can produce the exact response as the PLC for the same request, tested byte-by-byte. While \bca~is a straightforward way to assess LLM's accuracy, sometimes the bytes may not match for various reasons, for example, the physical process returns a different integer value despite the response packets being valid. Thus, a new metric, \rva, is defined to measure the ability of the LLM to respond with valid packets.

\begin{definition}[label={def:validation_accuracy}]{Response Validity Accuracy (\rva)}
A metric that measures if a produced response from the LLM is a valid network packet (see Table~\ref{table:modbus_s7-frame}) in terms of a protocol's specific fields, like \textit{transaction id} or \textit{function code}, that ignores the actual value of the analog or digital inputs/outputs and only validates if they are within the protocol defined range.
\end{definition}

Merely producing a valid network packet that is meaningless can be easily detected as a honeypot. Typically, attackers check if they are receiving a valid response and rarely further analyze the data fields of the response to understand whether the underlying physical process produces expected responses. For sophisticated attackers that would do further analysis,~\bca~needs to be maximized, as~\rva~maximization may not be sufficient. However, since the expected responses are based on the underlying process that could include noise and disturbances, they can potentially differ by some amount~$\epsilon$. This variation can still mislead the attacker into thinking a real process is being controlled. Therefore, we introduce a new metric~\rvae~that captures this ``epsilon'' difference.

\begin{definition}[label={def:pva_epsilon}]{\rva-Epsilon (\rvae)}
\rvae~expands the~\rva~metric by also checking the actual value of analog or digital inputs/outputs. The response is considered accurate if the value is within $\pm\epsilon$ of the actual value.
\end{definition}

The following equations stand given Definitions~\ref{def:accuracy}, ~\ref{def:validation_accuracy}, and~\ref{def:pva_epsilon}.
{\footnotesize
\begin{equation}
    \text{RVA-0}=\text{BCA}\,,\text{RVA-}\infty=\text{RVA}\,,\text{BCA} \leq \text{\rvae} \leq \text{RVA}
\end{equation}
}

\subsection{Base LLM}
\label{subsection:llm_exploration}

Generally, LLMs excel in generative tasks, primarily through a tokenizer, which converts incoming text into tokens that can then be used to formulate the response. Effective tokenization for network protocols would require a deep understanding of each protocol, considering the structured layers of packets, each with its unique fields and parameters. Thus, creating a universal tokenizer capable of handling multiple protocols poses significant challenges. A machine learning model capable of understanding network packets and generating responses without requiring tokenization is needed.

The ByT5 model~\cite{xue2022byt5}, developed by Google, introduces a novel approach to natural language processing by operating at the byte level, diverging from traditional token-based architectures. By processing text as a sequence of UTF-8 encoded bytes, ByT5 can handle a diverse range of languages and special characters without relying on a predefined vocabulary. This byte-level processing enables the model to manage inputs and outputs simply as byte sequences. Given this capability, if we consider only a specific ICS network protocol layer within the entire TCP packet stack, we can input just this segment to the ByT5 model and receive a response without needing to tokenize the packet. The \textit{finetuning process only needs two inputs, the source and target texts}. In the context of \shortacronym, this will correspond to the hexadecimal format of the request and response packets respectively. This method highlights the clear advantages of choosing the ByT5 model as the base LLM for \shortacronym, as it eliminates the need for tokenization.

%% file: sections/04-overview.tex
\begin{figure*}[!t]
\centering
\includegraphics[width=0.9\linewidth]{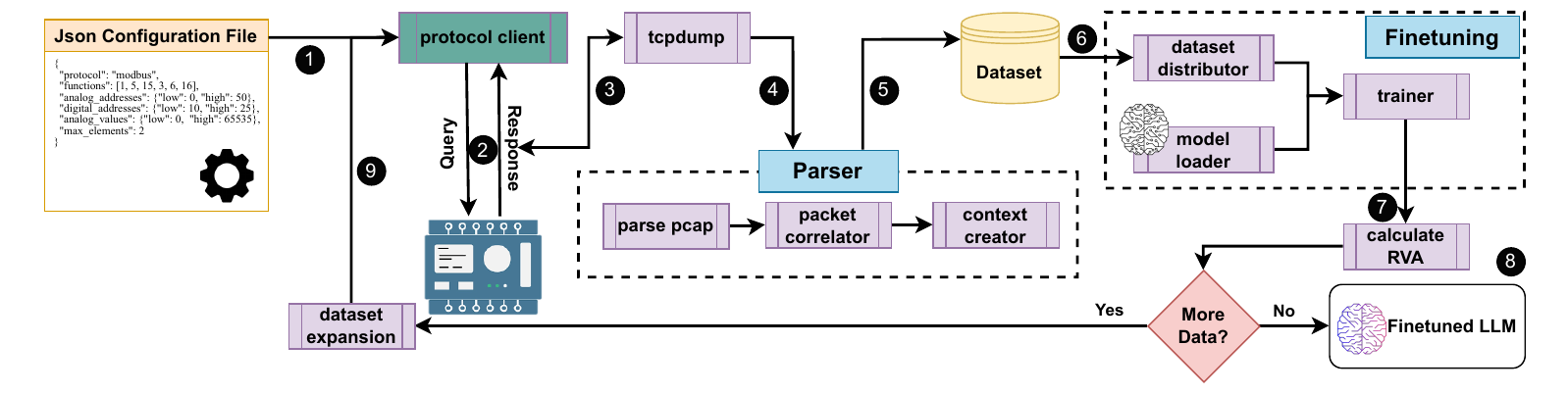}
\caption{\shortacronym's offline stage framework for dynamic configuration and PLC cloning. 
The process is fully automated, except for the box marked in the green (``protocol client''), which has to be performed manually for every new protocol. 
}
\label{fig:flow-diagram}
\end{figure*}

The basic components used in the offline process are described in the following subsections, while a graphical representation of them is depicted in Figure~\ref{fig:flow-diagram}.

\subsection{Dataset Generation Process}\label{subsection:dataset_generation}
The efficacy of fine-tuning LLMs for specific tasks heavily relies on the quality of the dataset used. Ensuring diversity across various scenarios is crucial in dataset generation, as it prevents model bias and enhances its ability to generalize across different conditions, leading to more robust performance.  

\subsubsection{PLC Profile}
Step~\blackcircle{1} in Figure~\ref{fig:flow-diagram}, illustrates the flow diagram of~\shortacronym's dataset generation process based on a user defined protocol specific configuration file. This file essentially describes the PLC device configuration and its parameters which are crucial for constructing a customized/user-specific protocol client.

\subsubsection{Protocol Client Initialization}
\textit{Protocol} attribute in the configuration file, specifies the communication layer that the PLC uses. The next attribute is the list of \textit{function} codes that are supported. Each function corresponds to a single or multiple read/write commands to a specific data type, digital or analog inputs/outputs, or device information. For example, the packet sample in Table~(\ref{table:modbus_s7-frame}, \blackcircle{a}) shows the FC as ``03'', which is decoded to read analog data from a specific address in Modbus, while FC ``04'' in Table~(\ref{table:modbus_s7-frame}, \blackcircle{c}) is decoded to read the data address from a specific data block area in S7comm, which are part of the official protocol documentation~\cite{modbus}. The ranges defined for \textit{analog\_addresses} and \textit{digital\_addresses} attributes determine the available accessible areas within the PLC memory.

\subsubsection{PLC Probing}
After the initialization, the protocol client establishes a connection to the target PLC device, allowing interaction through various request/response commands (step~\blackcircle{2}). Subsequently, an extensive network interaction traffic between the protocol client and the PLC is captured using a tool like Unix tcpdump~\cite{tcpdump} (step~\blackcircle{3}). The captured network traffic is stored in a packet capture (PCAP) file, which is sent as input to a parser (step~\blackcircle{4}) to extract the request and response commands. 

\subsubsection{PCAP Parser}
The parser is specifically designed to convert the raw binary data from the pcap file into hexadecimal byte strings. The packets might not be in sequence in the pcap file, thus the parser needs to correlate responses with their corresponding requests to ensure the correct interactive sequences. The data are saved in a CSV format file (step~\blackcircle{5}) that contains the request followed by the response.

\noindent\textbf{Context-Sensitive Dataset Generation:} For protocol emulation, historical data is not needed in the dataset. The dataset provides correct functionality effectively with individual requests and their corresponding responses since each pair is independent. However, when emulating physical processes, the response to a request might be affected by changes from previous interactions. Thus, for this purpose, the dataset must include contextual information to reflect the dynamic and interconnected nature of these events accurately. As a result, the PCAP parser includes an additional pre-processing step that organizes data into a context-aware format, where the dataset would look as follows:\\
{\small
\[req_{1}:resp_{1}|req_{2}:resp_{2}|{.}{.}|req_{n-1}:resp_{n-1}|req_{n}:,\,resp_{n}\]
}
where {\small $req_{i}$} and {\small $resp_{i}$} represent the $i$-th request and its corresponding response, respectively, and {\small$n \geq 2$}. This structured dataset helps in fine-tuning the LLM to produce context-aware responses and adjust to new queries based on past interactions.

\noindent\textbf{Automation:} Figure~\ref{fig:flow-diagram} highlights the module that needs manual effort with green color, ``protocol client" while the remaining modules are fully automated. The manual intervention in the ``protocol client" module is required in case of adapting new protocols which is further discussed in Section~\ref{subsection:dataset_generation_methodology}. Regarding the JSON configuration file, the user can configure it to match the desired PLC profile.

\subsection{LLM Fine-tuning Process}\label{subsection:llm_fine_tuning}
\shortacronym~utilizes the PyTorch Lightning framework~\cite{pytorch} to facilitate a structured method for model training, including efficient dataset management, validation during training, and post-training testing cycles. This framework has been adapted to fine-tune a pre-trained ByT5 model (e.g., google/byt5-small), which we use as our base LLM. Performance of the fine-tuned LLM is measured explicitly as how accurately the model produces replies to specific requests (see Definitions~\ref{def:accuracy},~\ref{def:validation_accuracy}, and~\ref{def:pva_epsilon}).

\subsubsection{Model Fine-tuning}\label{subsubsection:finetuning}
The adopted LLM fine-tuning process is depicted in step~\blackcircle{6}. The process begins with the dataset distributor module, which inputs the generated dataset into the trainer module. Simultaneously, the trainer module loads the pre-trained base LLM and initiates the fine-tuning phase. The trainer module performs a test cycle at the end of each training epoch, enabling continuous evaluation against a test dataset to monitor the actual model performance with an early stopping, where the fine-tuning phase is terminated if there is no improvement. The PCAP Parser dictates the value of ByT5 model's \textbf{\textit{source\_max\_token\_len}} and \textbf{\textit{target\_max\_token\_len}} hyper-parameters, which define the maximum input and output lengths respectively.

\subsubsection{Iterative Dataset Sizing}\label{subsubsection:iterative_dataset_sizing}
An appropriate dataset size selection is crucial when fine-tuning. A dataset that is too small may not fully capture the complexity of the task, leading to under-fitting. In contrast, an excessively large dataset can lead to over-fitting and unnecessary computational expenses. Therefore, the fine-tuning process begins with a small dataset size and iteratively doubles as the fine-tuned model is evaluated using the \rvae~metric (step~\blackcircle{7}). This is done through a dataset expansion module requesting additional data from the protocol client (step~\blackcircle{8}). The model is further fine-tuned with each increase in dataset size until no further improvements in the \rvae~metric are observed across multiple consecutive iterations. 

This approach provides a controlled and systematic exploration of the relationship between dataset size and model performance, identifying a point of diminishing returns where additional data no longer contributes to better model performance. The framework returns the fine-tuned LLM that effectively emulates the target PLC device.

%% file: sections/05-protocol_emulation.tex
Our initial goal is to fine-tune a pretrained LLM capable of accurately responding to network packets based on the network protocol used by the PLC, tailored to a specific device configuration. The dataset generation in this section will not encompass context, as historical data is not required for protocol emulation experiments.

\subsection{Dataset Generation Methodology}\label{subsection:dataset_generation_methodology}

To generate a representative dataset that emulates a specific network protocol, for example, Modbus protocol, an exhaustive dataset that consists of $65536^2$ (the maximum possible number of registers times the available values stored in 2 bytes per register) different requests is needed to cover all registers and their potential values for just one type of command (e.g., read or write). Following an exhaustive dataset generation strategy would demand significant resources for training. Therefore, Algorithm~\ref{alg:boundaries_client} implements the \textit{protocol\_client} shown in Figure~\ref{fig:flow-diagram}  which also describes the proposed \textit{boundaries} method. This method allows us to obtain a dataset that describes a protocol in a non-exhaustive manner while preserving accuracy.

By employing the \textit{boundaries} method, we significantly reduce the dataset size from a scale of billions to just a few thousand. This is achieved by covering the entire address and value space supported by a protocol with only selecting the minimum, maximum, and one randomly chosen intermediate value. 

The method starts with the \textit{execute} function~[line:~\ref{alg:fun:execute}], which essentially probes the PLC with the read/write commands included in the user defined \textit{functions} list [line: \ref{alg:function}]. The initial loop~[line:~\ref{alg:for:max_elements}] iterates through values from one to \textit{m\_elem} [line: \ref{alg:m_elem}], setting how many elements are going to be used from the selected read/write functions in the current loop. The second loop~[line:~\ref{alg:for:addresses}] iterates over values derived from the \textit{triplet} function~[line:~\ref{alg:fun:triplet}]. This function generates three specific values: the lowest, the highest minus the provided elements, and a random number in between, which forms the basis of the \textit{boundaries} method. To add a layer of realism, this method includes requests that trigger erroneous responses (i.e., exceptions). Exceptions can occur when a request is made using an inactive address outside the valid range. This approach allows the dataset to reflect typical operational data and include scenarios that simulate potential error conditions.

\begin{algorithm}[!t]
\caption{\small \shortacronym~Protocol Emulation using \textit{boundaries}}
\label{alg:boundaries_client}
\begin{algorithmic}[1]
\footnotesize
\Require~~\\
$read,\,write$: abstract functions specific to the network protocol.\\
$m\_elem$ (max\_elements): the number of elements up to which to read or write using only one command.\label{alg:m_elem}\\
$a\_low,\,a\_high$: the range of addresses available according to the protocol running on the PLC.\\
$max\_addr$: max address supported by PLC-protocol combination.\\
$val\_low,\,val\_high$: the range of values that the PLC is configured to reply with normal responses.\\
$bounds$: the range for normal operations\\
$e\_bounds$: the range of values for exception cases (address out of bounds)\\
$functions$: the list of protocol specific functions to call.\\

\Function{execute}{}\label{alg:fun:execute}
  \For{$elem$ \textbf{in} $range(1,\,m\_elem)$}\label{alg:for:max_elements}
    \State $bounds \gets \Call{triplet}{a\_low, a\_high, m\_elem}$
    \State $e\_bounds \gets \Call{triplet}{a\_high + 1, max\_addr, m\_elem}$\label{alg:exceptions}
    \For{$addr$ \textbf{in} $[bounds, e\_bounds]$}\label{alg:for:addresses}
      \State $cartesian \gets \Call{combs}{val\_low,\,val\_high, elem}$
      \For{$data$ \textbf{in} $cartesian$}\label{alg:for:data}
        \For{FUNC \textbf{in} $functions$}\label{alg:for:functions}
            \State $\Call{func}{addr\,\,data,\,elem}$\label{alg:function}
        \EndFor
      \EndFor
    \EndFor
  \EndFor
\EndFunction
\\
\Function{combs}{$low,\,high,\,elem$}\label{alg:fun:combs}
    \State $bounds \gets \Call{triplet}{low, high, 0}$
    \State $combs \gets \Call{cartesian\_product}{bounds,\,elem}$\label{alg:fun:cartesian}
    \State $result \gets \text{empty dictionary}$
    \For{$comb,\,index$ \textbf{in} $combs$}
        \State $result[index] \gets comb$
    \EndFor
    \State \Return $result$
\EndFunction
\\
\Function{triplet}{$low,\,high,\,elem$}\label{alg:fun:triple}\label{alg:fun:triplet}
    \State \Return $[low,\,\Call{rand}{low+1,\,high-elem-1},\,high-elem]$
\EndFunction
\end{algorithmic}
\end{algorithm}

Before the loop~[line:~\ref{alg:for:data}], the \textit{combs} function~[line:~\ref{alg:fun:combs}] is called to produce all possible data combinations based on the given range of possible values and the number of elements to accommodate. \textit{Combs} function starts by calling the \textit{triplet} [line:~\ref{alg:fun:triplet}] function to obtain a set of three numbers. These numbers are used to compute the Cartesian product of the specified number of elements [line:~\ref{alg:fun:cartesian}], which is later used in the request commands that include multiple inputs. By that, all possible combinations of values for the number of elements are covered. The resulting values are formatted into triplets to be used in the \textit{execute} function. The \textit{execute} function then sends requests to a PLC or simulator, which is called repeatedly to align with the dataset size in the configuration file.

\textbf{Why boundaries method} This algorithm significantly enhances the adaptability, scalability, and efficiency of ICS honeypots by dynamically generating training data across a range of operational states. This method ensures a smart and effective dataset generation for training the LLM to understand the range of accessible addresses/values without the need to generate traffic that includes them. This helps in making the framework more dynamic and adaptable to various PLC configurations, moving away from the traditional reliance on static and manual scripting specific to PLC setups.

\textbf{Adapting New Protocols} Algorithm~\ref{alg:boundaries_client} serves as the foundation for integrating a ``protocol client" module (Figure~\ref{fig:flow-diagram}) for other PLC protocols. Step 1: Identify the terminology used in the new protocol and map it to the corresponding terms in the existing implementations. For example, Modbus uses ``registers" for analog input/output, while S7comm employs ``datablocks". Step 2: Identify the valid addresses and value ranges of the new protocol. Step 3: Identify the list of available function codes. Step 4: Modify the default configuration file to include the identified addresses and values into the analog inputs/outputs. Step 5: Instantiate Algorithm~\ref{alg:boundaries_client} per our examples existing in our codebase using readily available libraries to generate the protocol client module. The rest of the methodology is automated.

\subsection{Fine-tuning LLM}

\subsubsection{Ablation Study on input/output configuration}\label{subsubsection:cfg_ablation}
\begin{figure}[!t]
\subfloat[\bca~on Modbus protocol]{\includegraphics[width=0.49\linewidth]{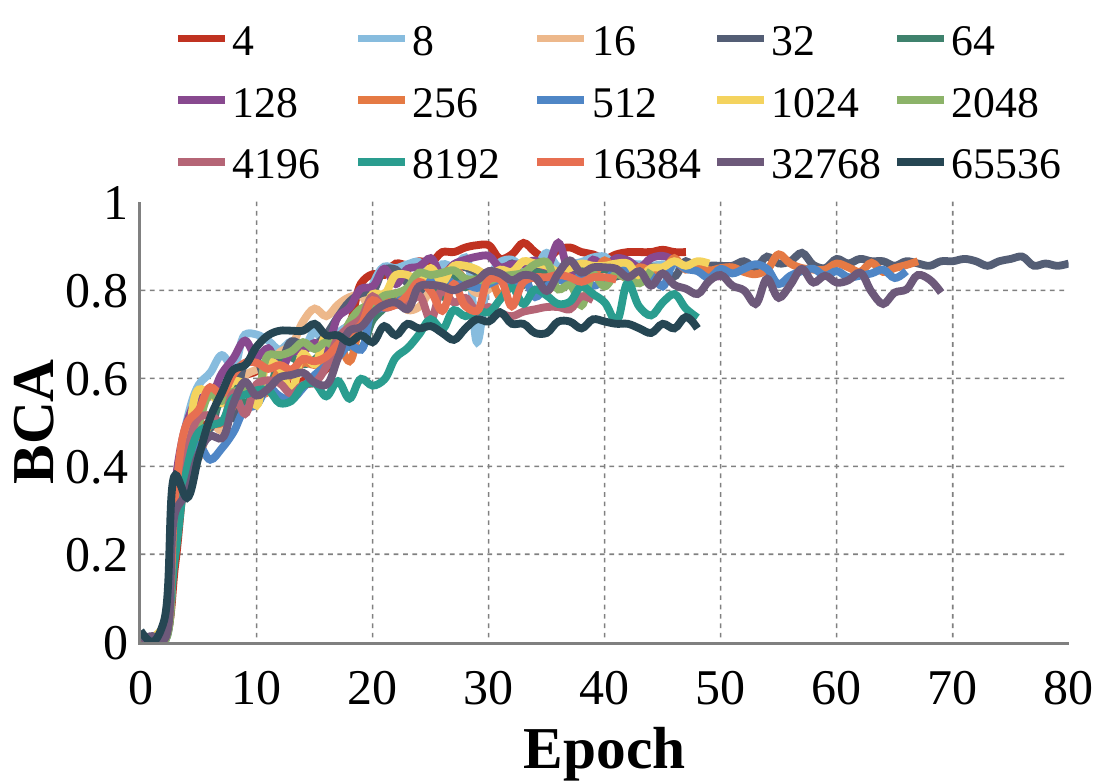}}\label{subfig:ablation:protocol_input_output:mbtcp:exact}
\subfloat[\rva~on Modbus protocol]{\includegraphics[width=0.49\linewidth]{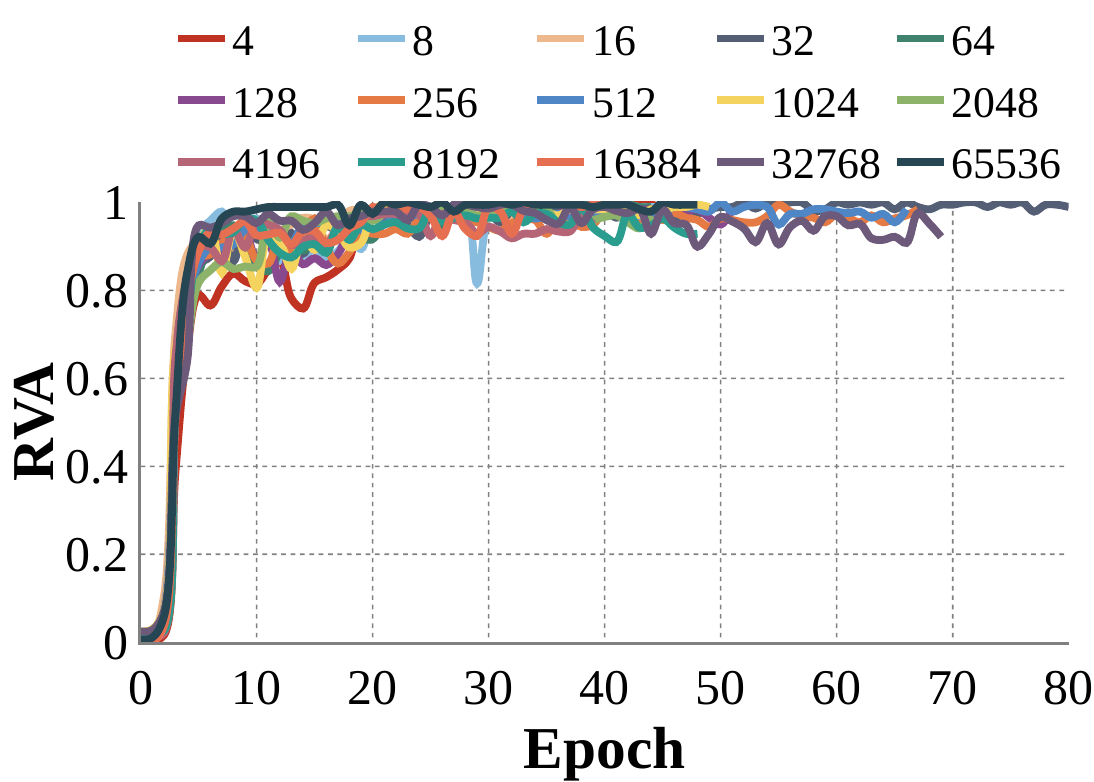}}\label{subfig:ablation:protocol_input_output:mbtcp:validator}
\caption{\bca~and~\rva~per epoch of the byt5-small model when using different input/output configurations ($2^2$: 4 up to $2^{16}$: 65536). Similar values on the metrics prove that~\shortacronym~is independent of the PLC configuration.}
\label{fig:ablation:protocol_input_output:performance}
\end{figure}

To prove~\shortacronym's independence of different input/output configurations, an ablation study was performed experimenting with powers of 2 to cover the maximum available inputs/outputs, 65536 which is $2^{16}$. Given so, Figure~\ref{fig:ablation:protocol_input_output:performance} presents the experiment starting with $2^2$ up until $2^{16}$, excluding $2^0$ and $2^1$ as boundaries cannot be applied in this case, preventing any significant dataset minimization. This experiment further demonstrates that choosing any input/output configuration does not impact the \bca~and~\rva~metrics or the overall performance of the LLM on the selected protocol. Both metrics exhibit the same trend, with variations attributed solely to the stochastic nature of LLMs. The experiments performed in the following sections use a setup that is attributed to specific hardware-in-the-loop testbeds employed at our lab.

\subsubsection{Ablation Study on model and dataset sizes}\label{subsubsection:ablation_size}
We run the protocol emulation experiments for byt5-small, byt5-base, and Llama3.2~\cite{touvron2023llama} which have 300 million, 580 million, and 1 billion parameters respectively. Table~\ref{table:protocol_emulation:model_sizes} depicts that there is no advantage to using the bigger model (byt5-base), especially for dataset sizes above 1600 samples, as the results are relatively close. As the dataset size increases above 3200 samples, overfitting is experienced and thus we see a drop in performance of both byt5 model versions. The fact that byt5-small performs marginally worse in the 6400 samples dataset in terms of~\bca~and~\rva~is not particularly significant due to the impact of overfitting. The main objective of~\shortacronym~is to fine-tune the LLM with a small dataset that gives us good metric values due to the boundaries method.

On the other hand, Llama3.2 \bca~and \rva~metrics were almost zero with very few correct emulated packets. This can be attributed to the tokenization need for this type of model. Finally, we use LoRA~\cite{hu2021lora} to minimize training time on the byt5-small model by training 40 million instead of 300 million parameters, but this also yields poor results which can be attributed to the fact that we re-purpose the model's objective in generating hexadecimal responses instead of doing language translation.

\begin{takeaway}
There is no need to move to a bigger model since similar results, on the~\rva~metric, can be achieved using different dataset sizes and no substantial advantage will be gained by the additional parameters of a bigger model. Hence, we consider byt5-small for the remaining experiments.
\end{takeaway}

\begin{table}[!t]
    \centering
    \caption{Ablation study on the size of ByT5 model.}
    \label{table:protocol_emulation:model_sizes}
    \begin{tabularx}{\columnwidth}{|c|c|Y|Y|}
         \hline
         \thead{\textbf{Size}} & \thead{\textbf{Model}} & \thead{\textbf{\bca}} & \thead{\textbf{\rva}} \\
         \hline
         \multirow{2}{*}{200}  & small & $59.50 \pm 6.35$ & $85.00 \pm 7.53$ \\
                               \cline{2-4}
                               & base & $71.13 \pm 5.29$ & $96.43 \pm 3.96$ \\
         \hhline{|=|=|=|=|}
         \multirow{2}{*}{400}  & small & $71.94 \pm 8.27$ & $93.33 \pm 5.59$ \\
                               \cline{2-4}
                               & base & $73.04 \pm 7.92$ & $91.25 \pm 8.13$ \\
         \hhline{|=|=|=|=|}
         \multirow{2}{*}{800}  & small & $85.83 \pm 0.88$ & $98.47 \pm 1.95$ \\
                               \cline{2-4}
                               & base & $85.00 \pm 5.96$ & $96.88 \pm 6.14$ \\
         \hhline{|=|=|=|=|}
         \multirow{2}{*}{1600} & small & $85.31 \pm 1.25$ & $99.30 \pm 0.62$ \\
                               \cline{2-4}
                               & base & $82.55 \pm 3.97$ & $95.87 \pm 4.79$ \\
         \hhline{|=|=|=|=|}
         \multirow{2}{*}{3200} & small & $80.27 \pm 0.97$ & $97.77 \pm 1.42$ \\
                               \cline{2-4}
                               & base & $79.67 \pm 2.91$ & $97.34 \pm 2.83$ \\
         \hhline{|=|=|=|=|}
         \multirow{2}{*}{*6400} & small & $82.62 \pm 1.49$ & $98.81 \pm 1.10$ \\
                               \cline{2-4}
                               & base & $82.74 \pm 2.10$ & $98.91 \pm 1.31$ \\
         \hline
    \end{tabularx}
    \vspace{0.1cm}
    \par *Dataset sizes above 3200 samples experience overfitting.\\
\end{table}

\begin{figure}[!t]
\centering
\includegraphics[width=\linewidth]{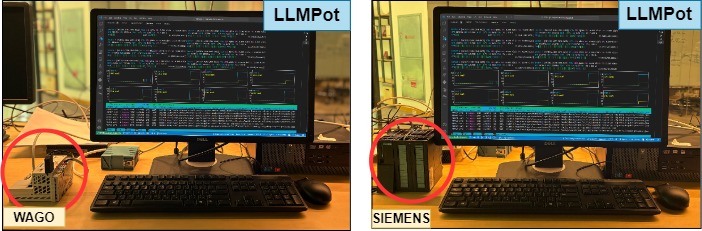}
\caption{PLCs used in our experiments, WAGO:Modbus\cite{wago} (left) and SIEMENS:S7comm\cite{siemens} (right), being copied by \shortacronym.}
\label{fig:plc_devices}
\end{figure}

\subsubsection{Protocol Generalization}
To demonstrate~\shortacronym's generalizability in emulating different protocols, we evaluated \shortacronym~on WAGO and SIEMENS PLC devices (see Figure~\ref{fig:plc_devices}), which use Modbus and S7comm protocols, respectively. Following Algorithm~\ref{alg:boundaries_client}, our exploration starts by generating several combinations of dataset sizes as shown in Configuration~\ref{def:datasets} to identify the optimal dataset size and obtain the best performance. Based on the minimum number of samples generated in one iteration by Algorithm~\ref{alg:boundaries_client}, which is 144 samples, the minimum size of the fine-tuning dataset is determined irrespective of the protocol and is dictated by the \textit{max\_elements} parameter specified in the configuration file. Given that, the minimum meaningful sample size is set to 200 and doubled until no further improvement is observed, as discussed in Section~\ref{subsubsection:iterative_dataset_sizing}. 

\begin{figure}[!t]
\vspace{-0.5em}
\subfloat[\bca~on Modbus protocol]{\includegraphics[width=0.49\linewidth]{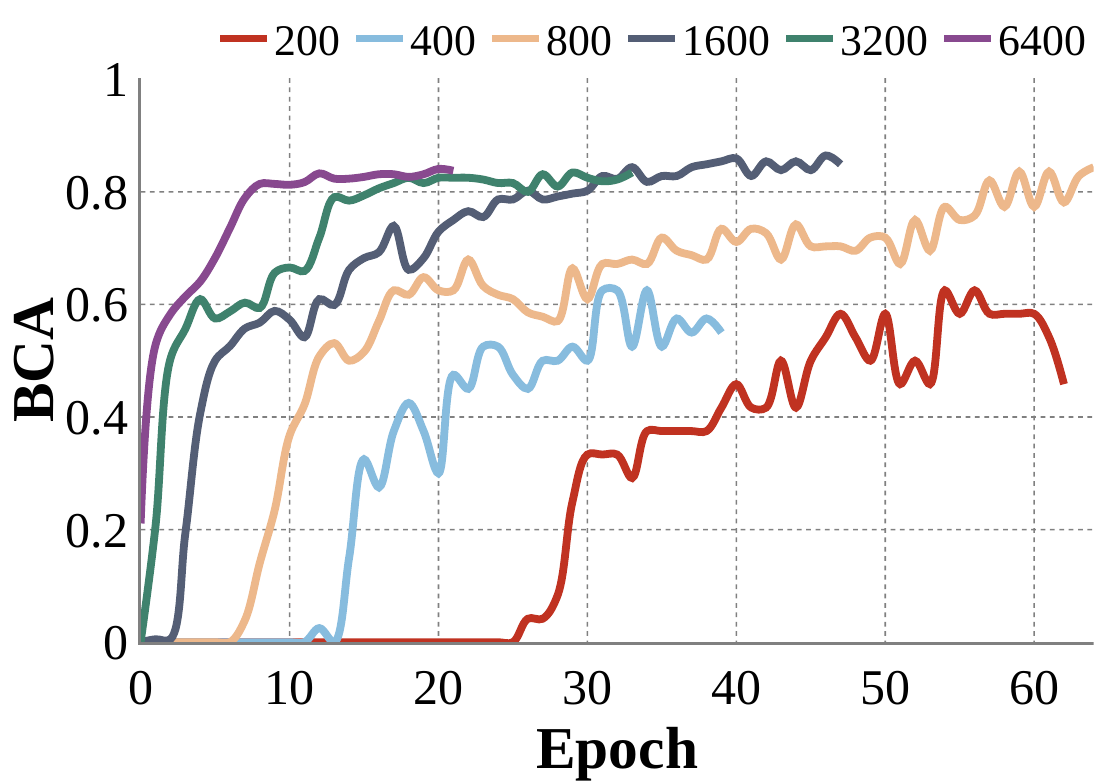}}\label{subfig:protocol_emulation:mbtcp:exact}
\subfloat[\rva~on Modbus protocol]{\includegraphics[width=0.49\linewidth]{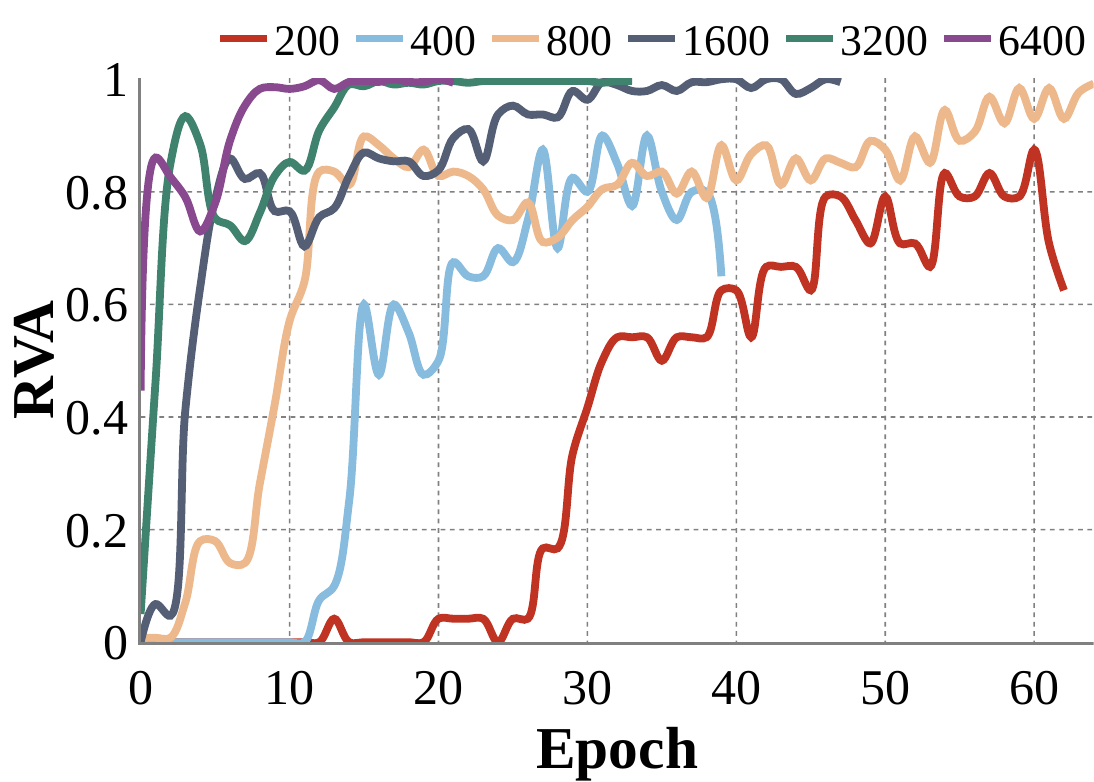}}\label{subfig:protocol_emulation:mbtcp:validator}
\hfill
\subfloat[\bca~on S7comm protocol]{\includegraphics[width=0.5\linewidth]{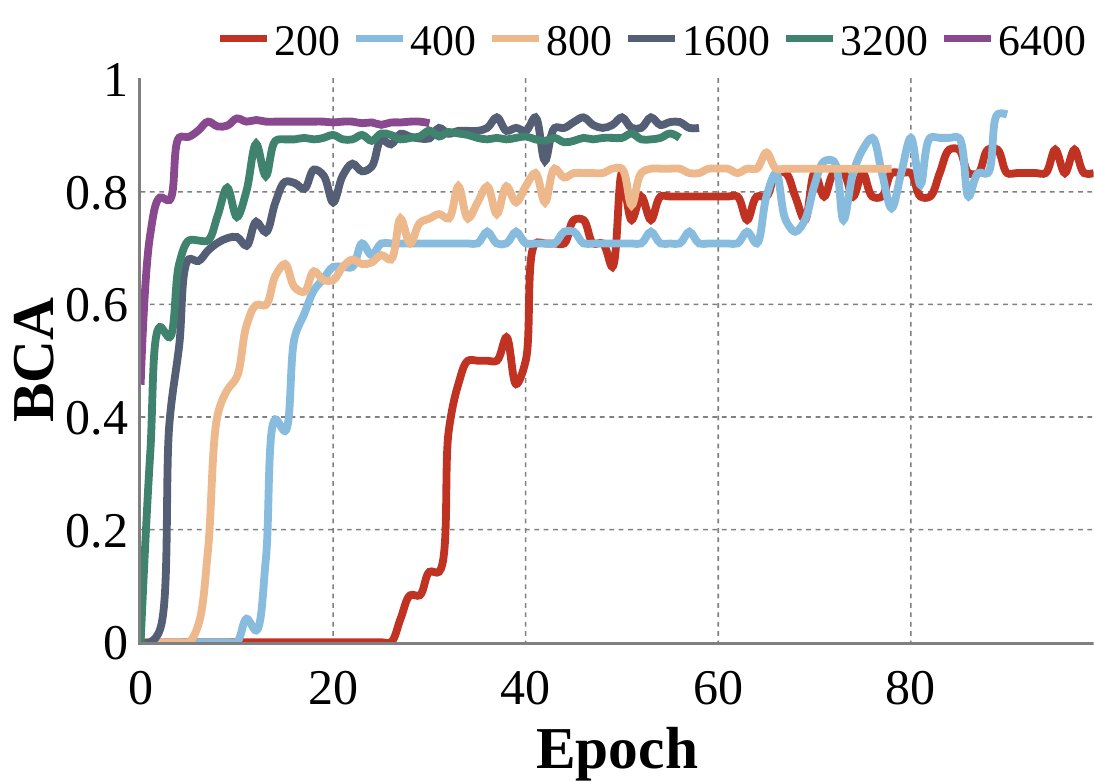}\label{subfig:protocol_emulation:s7comm:exact}}
\subfloat[\rva~on S7comm protocol]{\includegraphics[width=0.5\linewidth]{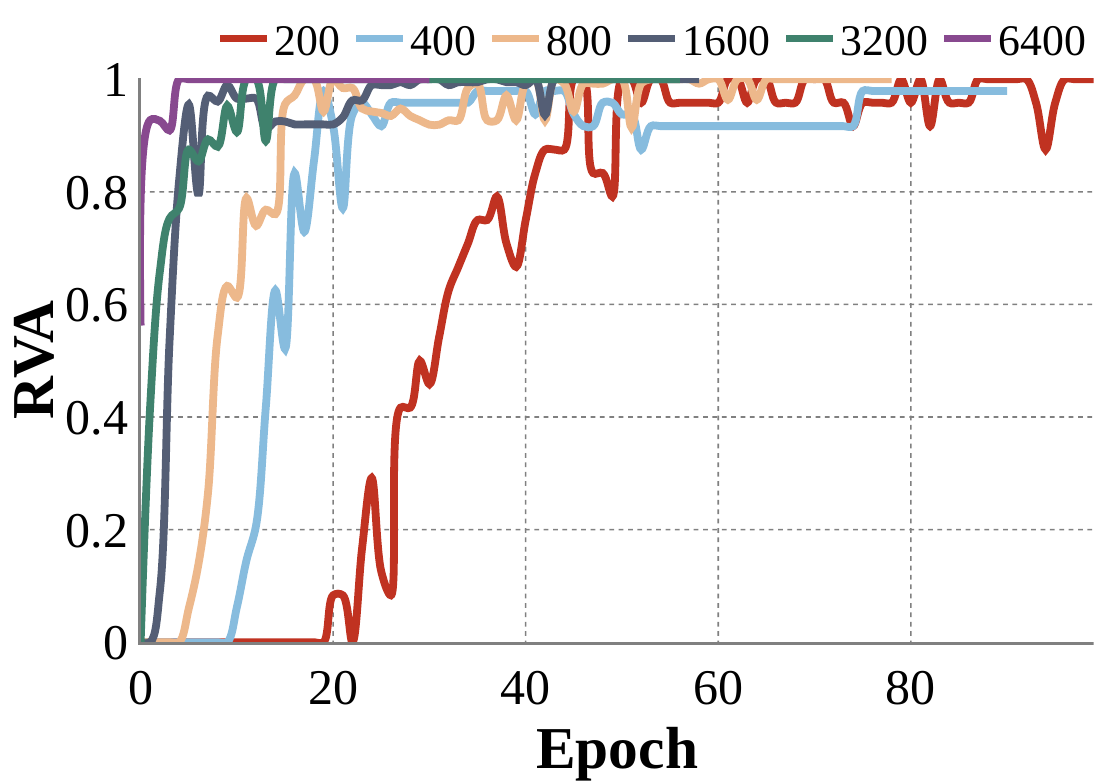}\label{subfig:protocol_emulation:s7comm:validator}}
\caption{\bca~and~\rva~per epoch of the byt5-small model when using different dataset sizes and protocols to fine-tune. A patience value of 10 epochs was used to stop the fine-tuning in case the validation loss was not improving, thus the different ending epochs for each dataset size.}
\label{fig:protocol_emulation:performance}
\end{figure}

Moreover, we focus on a specific set of read and write commands that handle both analog and digital inputs/outputs since these are involved in physical processes. As shown in Configuration~\ref{def:datasets}, the covered subset of functions are denoted by their particular numerical FCs (digital read/write: 1, 5, 15, and analog read/write: 3, 6, 16). It is worth pointing out that the LLM fine-tuned only on the functions specified in the configuration file will respond with an invalid packet if it encounters any other function. Furthermore, the range of the number of permissible inputs and outputs across the allowable range for both Modbus and S7comm protocol was randomly chosen to be 40 for both analog and digital inputs/outputs.

\begin{configuration}[label={def:datasets}]{Fine-tuning Datasets}
    \textbf{Model}: byt5-small\\
    \textbf{Protocols}: Modbus, S7comm\\
    \textbf{Sizes}: 200, 400, 800 1600, 3200, 6400 samples\\
    \textbf{Split}: 80\% training, 10\% validation, 10\% testing\\
    \textbf{Functions*}: 1, 5, 15, 3, 6, 16\\
    \textbf{Digital*}: 40\\
    \textbf{Analog*}: 40
    \footnotesize
    \par Each dataset essentially corresponds to a separate fine-tuned model.
    \par The trained model with dataset size 1600 is marked as \blackcircle{o1}.
    \par *Generalizability is tested and proven in Section~\ref{subsubsection:cfg_ablation}
\end{configuration}

During the fine-tuning phase, we evaluate the LLM's performance using the \bca~and \rva~metrics at the end of each epoch, using 10\% of the fine-tuning dataset. This assessment helps determine how many epochs are necessary for the model to reach its maximum performance, given the dataset size. Figure~\ref{fig:protocol_emulation:performance} presents the \bca~and the \rva~for both protocols for different dataset sizes. We can observe that the \rva~improves more rapidly and achieves higher values as the dataset size increases. For smaller datasets, overfitting is evident as depicted by the validation loss surpassing the training loss (see Figure~\ref{subfig:appendix:protocol_emulation:mbtcp:loss} and Figure~\ref{subfig:appendix:protocol_emulation:s7comm:loss} in Appendix~\ref{appendix:protocol_emulation}). Additionally, losses for these smaller datasets remain relatively high, indicating inadequate fine-tuning. 

\begin{takeaway}\label{takeaway:finetuning_results}
Byt5-small can effectively learn and respond to network protocols, providing responses with high \rva, and verifying the responses are valid network packets.
\end{takeaway}

\begin{takeaway}\label{takeaway:finetuning_saturation}
Saturation of the model's performance with increasing dataset size supports the effectiveness of using the \textit{boundaries} method to create a smaller fine-tuning dataset rather than adopting an exhaustive approach for dataset generation.
\end{takeaway}

\subsubsection{Experimental Evaluation}\label{subsection:experimental-evaluation}

We also experimented with different numbers of permissible inputs and outputs across the allowable ranges for both the Modbus and S7comm protocols, testing values from smaller sets (e.g., 40, 100) to larger ones (e.g., 5000, 10000), refer to Table~\ref{table:protocol_emulation:test-datasets}. This aimed to demonstrate the efficiency of protocol emulation across varying input/output sizes by using 1600 samples as the dataset size, based on the findings from Section~\ref{subsubsection:ablation_size}.

Based on that, we demonstrate~\shortacronym's generalizability with two datasets, \blackcircle{g1} and \blackcircle{g2}, each with 1600 samples but differing in the number of active addresses for inputs/outputs, 100 and 5000 respectively. Additionally, we consider two more distinct emulation datasets: \blackcircle{d1}, having the same PLC configurations as the fine-tuning datasets, and \blackcircle{d2}, which utilizes different configurations (10000 instead of 40 inputs/outputs). 

\begin{table}[!t]
    \caption{Additional datasets used to fine-tune ByT5 model and run experimental evaluations. Each dataset essentially represents a fine-tuned model.}
    \label{table:protocol_emulation:test-datasets}
    \centering
    \begin{tabular}{ |c|c|c|c|c| }
        \hline
        \multirow{2}{*}{\thead{\textbf{Characteristics}}} & \multicolumn{2}{c|}{\thead{\textbf{Generalization}}} & \multicolumn{2}{c|}{\thead{\textbf{Emulation}}} \\
        \cline{2-5}
        & \thead{\blackcircle{g1}} & \thead{\blackcircle{g2}} & \thead{\blackcircle{d1}} & \thead{\blackcircle{d2}}\\
        \hline
        Sizes & \multicolumn{4}{c|}{1600} \\
        \hline
        Functions & \multicolumn{4}{c|}{1-5-15-3-6-16} \\
        \hline
        Number of Digital & 100 & 5000 & 40 & 10000 \\
        \hline
        Number of Analog & 100 & 5000 & 40 & 10000 \\
        \hline
    \end{tabular}
\end{table}

\subsubsection*{Address Range Generalization}
\begin{figure}[!t]
\vspace{-1.1em}
\subfloat[\bca~ on Modbus protocol]{\includegraphics[width=0.49\linewidth]{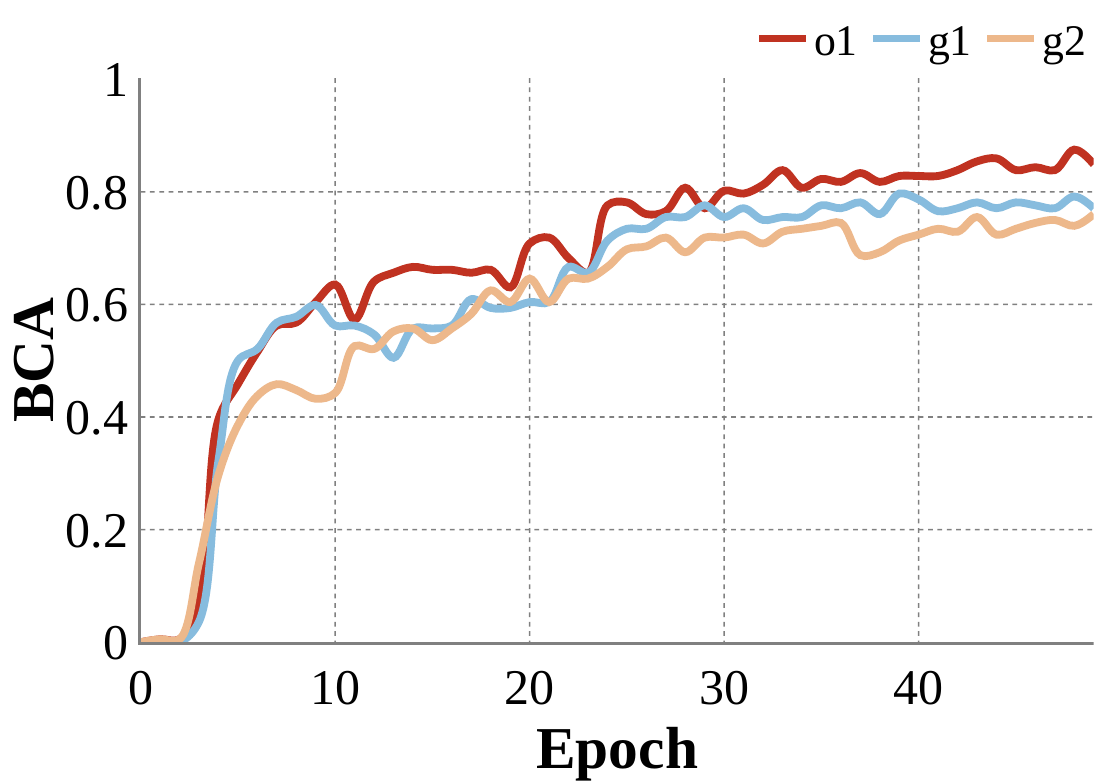}\label{subfig:ablation:mbtcp:bca}}
\subfloat[\rva~on Modbus protocol]{\includegraphics[width=0.49\linewidth]{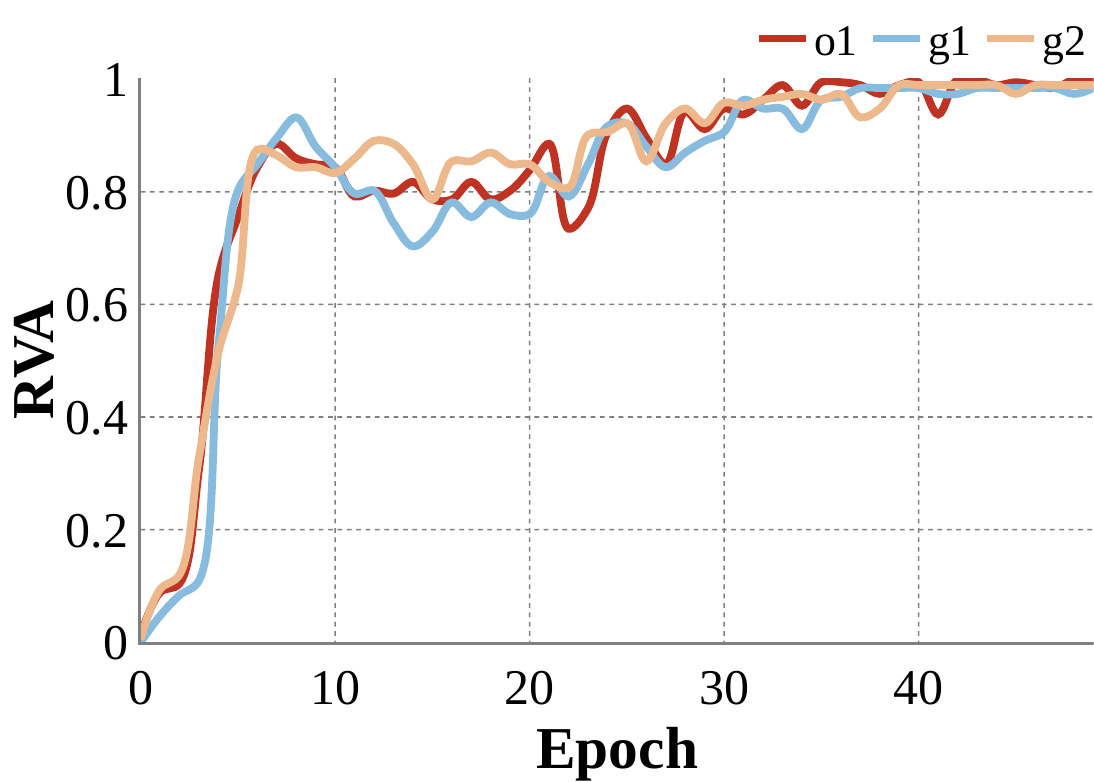}\label{subfig:ablation:mbtcp:pva}}

\subfloat[\bca~on S7comm protocol]{\includegraphics[width=0.5\linewidth]{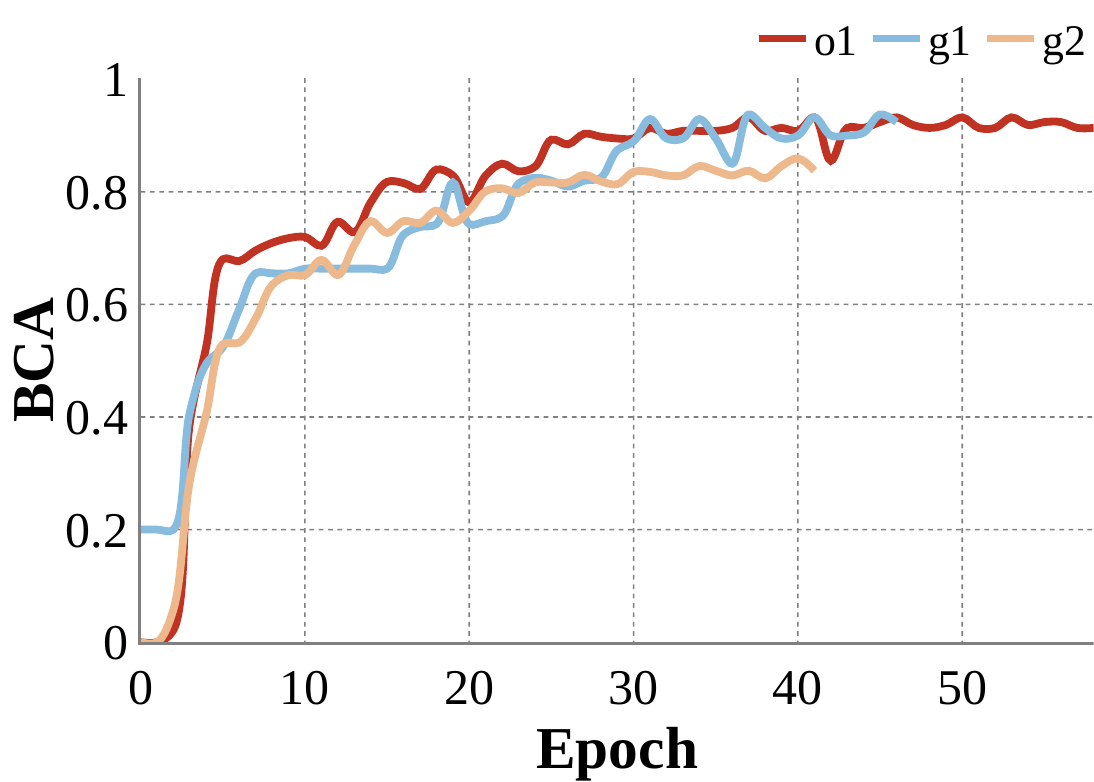}\label{subfig:ablation:s7comm:bca}}
\subfloat[\rva~on S7comm protocol]{\includegraphics[width=0.5\linewidth]{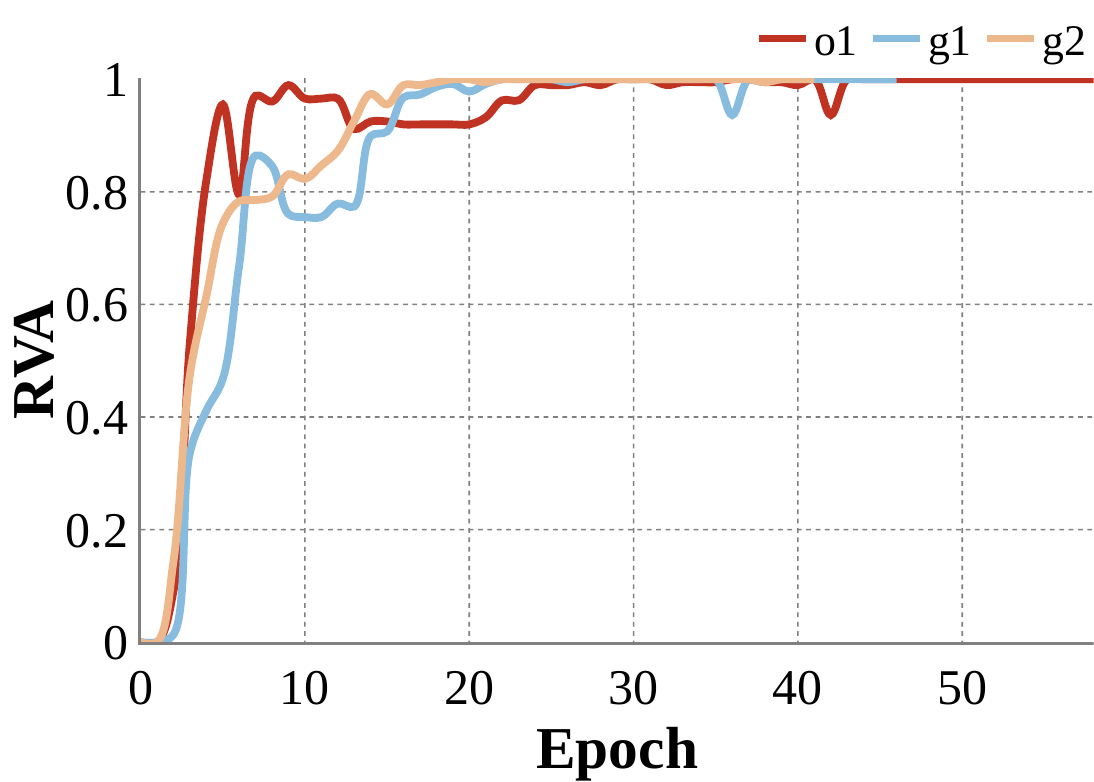}\label{subfig:ablation:s7comm:pva}}

\caption{\bca~and \rva~per epoch of the byt5-small model when using different configurations of analog and digital inputs/outputs \protect\blackcircle{o1}, \protect\blackcircle{g1}, \protect\blackcircle{g2}.}
\label{fig:ablation:mbtcp:bca-pva}
\end{figure}

We use models \blackcircle{o1}, \blackcircle{g1} and \blackcircle{g2} to demonstrate the generalizability of \shortacronym~across varying numbers of active addresses for digital and analog inputs/outputs in PLC setups. Figure~\ref{subfig:ablation:mbtcp:bca} and Figure~\ref{subfig:ablation:s7comm:bca} present the \bca~and Figure~\ref{subfig:ablation:mbtcp:pva} and Figure~\ref{subfig:ablation:s7comm:pva} present the \rva~metrics for different combinations of analog and digital inputs/outputs for both Modbus and S7comm protocols. We can observe that all the plots share similar characteristics with almost identical final values. 
The \bca~values over the training epochs converge closely, highlighting the stochastic nature of LLM training. Whereas the \rva~results present similar outcomes but with slight improvement due to the metric's characteristics. The dataset generation here is balanced and not biased toward any particular function to ensure fair representation.

\begin{takeaway}
Regardless of the PLC configuration in terms of address availability for digital and analog inputs/outputs, the model’s performance remains consistent when the same dataset size is used for fine-tuning.
\end{takeaway}

\subsubsection*{PLC profiling given different digital/analog configurations}
Given datasets \blackcircle{d1} and \blackcircle{d2} shown in Table~\ref{table:protocol_emulation:test-datasets}, we fine-tuned byt5-small, essentially profiling two PLCs with different digital/analog configurations. \textit{Do these two models properly reply to corner case requests?} 

Table~\ref{table:protocol_emulation:example} depicts the corner case requests, the ground truth results for both PLC configurations \blackcircle{d1} and \blackcircle{d2}, the response from the model fine-tuned based on Configuration~\ref{def:datasets}, and the corresponding \bca~and \rva~metrics result. The model's responses in bold mean a correct response and striked-out ones mean an unexpected or erroneous response. Next columns mark if the reply of the model is correct (\cmark) or wrong (\xmark) with respect to \bca~and \rva~metrics.

\begin{table}[!t]
  \caption{Example request-response set for the different emulation scenarios, the model's responses and the validation using both~\bca~and~\rva.}
  \label{table:protocol_emulation:example}
  \begin{tabularx}{\columnwidth}{ |c|Y|Y|c|Y|Y|Y|Y| }
    \hline
    \multirow{2}{*}{\thead{\textbf{Request}}} & \multicolumn{2}{c|}{\textbf{\thead{Ground\\Truth}}} & \multirow{2}{*}{\rotatebox[origin=c]{90}{\textbf{\thead{Model\\Response}}}} & \multicolumn{2}{c|}{\textbf{\thead{\bca}}} & \multicolumn{2}{c|}{\textbf{\thead{\rva}}} \\
    \cline{2-3}\cline{5-8}
     & \raisebox{-0.5ex}{\thead{\blackcircle{d1}}} & \raisebox{-0.5ex}{\thead{\blackcircle{d2}}} &  & \raisebox{-0.5ex}{\thead{\blackcircle{d1}}} & \raisebox{-0.5ex}{\thead{\blackcircle{d2}}} & \raisebox{-0.5ex}{\thead{\blackcircle{d1}}} & \raisebox{-0.5ex}{\thead{\blackcircle{d2}}}\\
    \hline
    read(addr=41) & exc & val & \textbf{exc} & \cmark & \xmark & \cmark & \cmark\\
    \hline
    read(addr=41) & exc & val & \sout{val} & \xmark & \cmark & \xmark & \xmark\\
    \hline
    read(addr=10001) & exc & exc & \textbf{exc} & \cmark & \cmark & \cmark & \cmark\\
    \hline
    read(addr=10001) & exc & exc & \sout{val} & \xmark & \xmark & \xmark & \xmark\\
    \hline
    read(addr=35) & val & val & \sout{exc} & \xmark & \xmark & \xmark & \xmark\\
    \hline
    read(addr=35) & val & val & \textbf{val} & \cmark & \cmark & \cmark & \cmark\\
    \hline
  \end{tabularx}
\end{table}

As an example, in the second line of Table~\ref{table:protocol_emulation:example}, given the request to read address 41, the reply in \blackcircle{d1} dataset is an exception since the valid address range was up to 40, while in dataset~\blackcircle{d2} includes a reply with a value of this address because the valid range is up to 10,000. Considering that we use~\blackcircle{d1} as a dataset for the validator, \bca~will mark this as a wrong answer but correct if we use \blackcircle{d2}. It is worth mentioning here that \bca~is just the exact match of the model's reply against the one in the dataset. For \rva, in both cases, the reply is wrong, because the validator checks, given the requested address what should have been the correct reply, considering the PLC configuration that was used to train the model.

\begin{figure}[!t]
\vspace{-1.2em}
\subfloat[\bca~on Modbus protocol]{\includegraphics[width=0.5\linewidth]{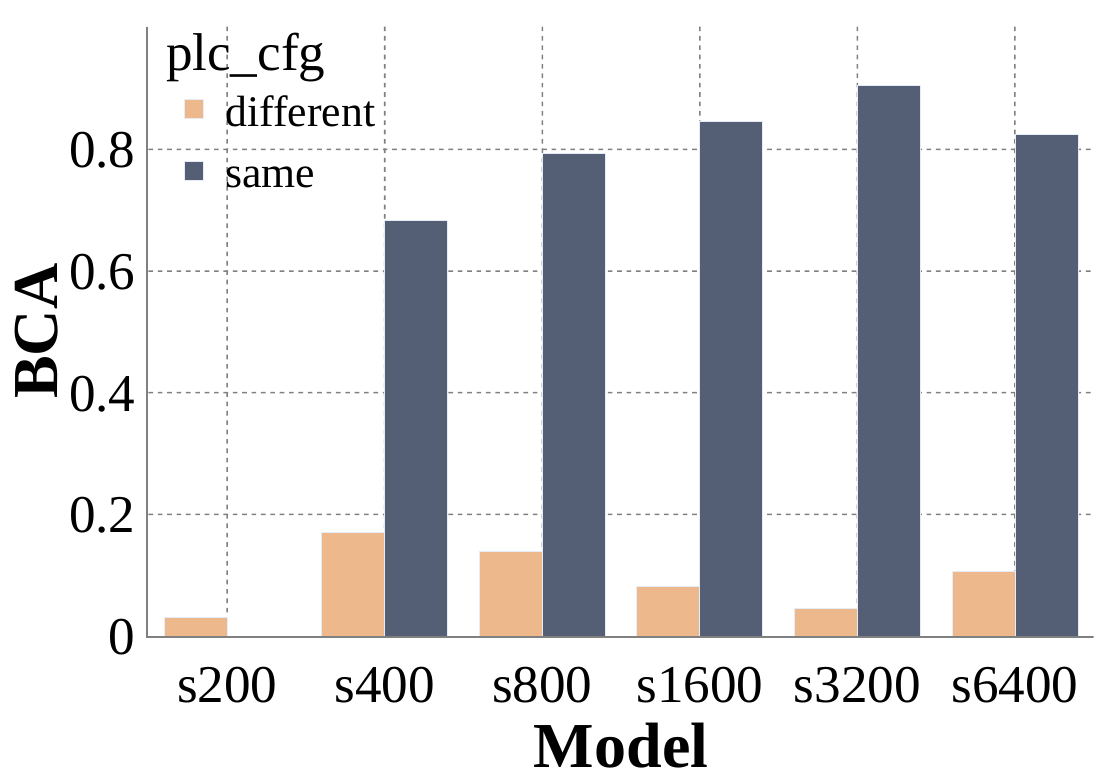}\label{subfig:protocol_emulation:exact:emulation_dataset}}
\subfloat[\rva~on Modbus protocol]{\includegraphics[width=0.5\linewidth]{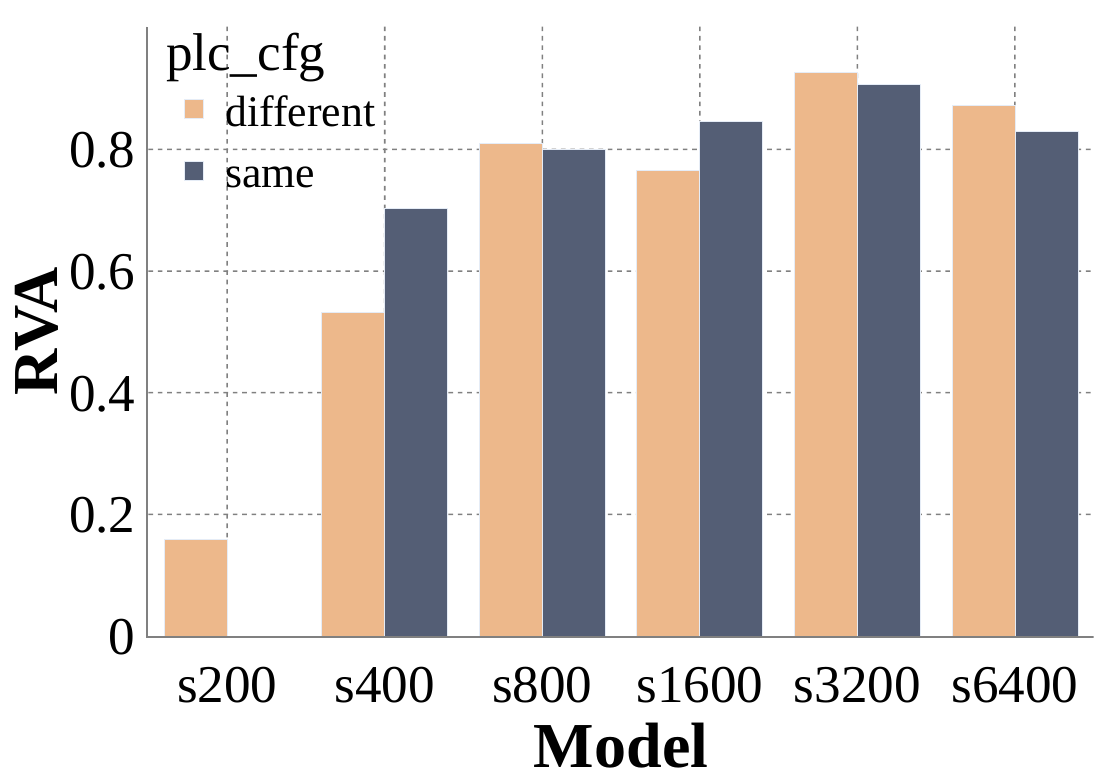}\label{subfig:protocol_emulation:validator:emulation_dataset}}
\hfill
\subfloat[\bca~on S7comm protocol]{\includegraphics[width=0.5\linewidth]{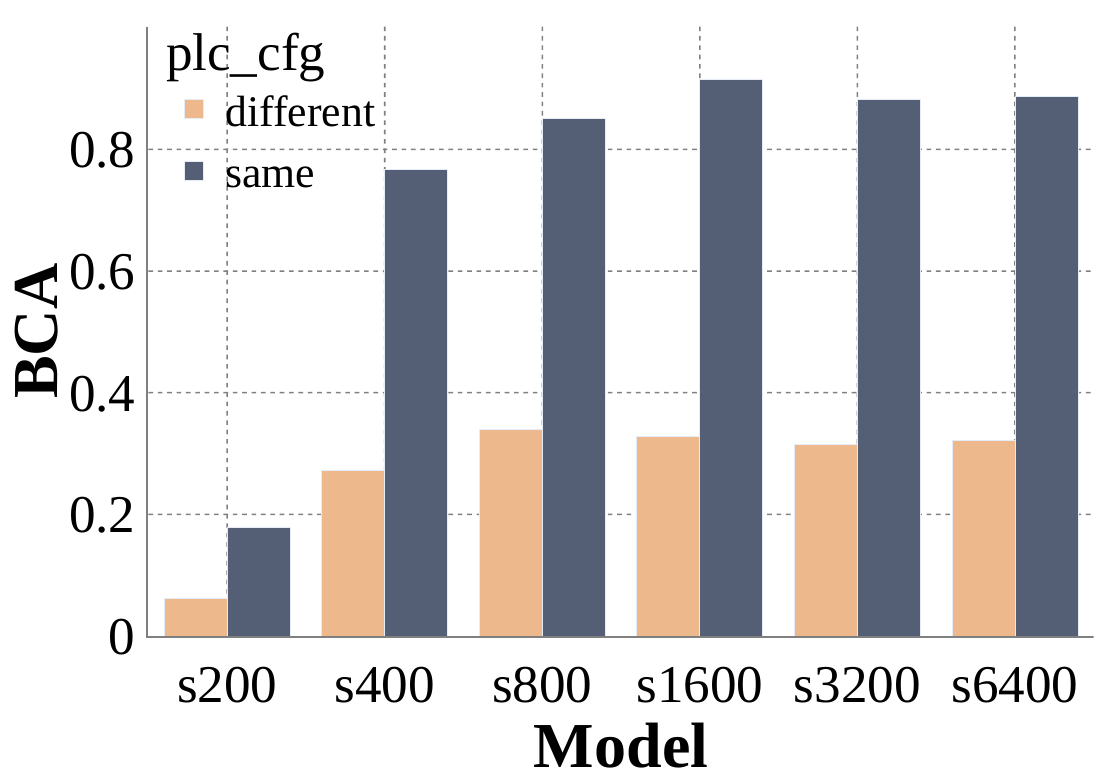}\label{subfig:protocol_emulation:s7comm:exact:emulation_dataset}}
\subfloat[\rva~on S7comm protocol]{\includegraphics[width=0.5\linewidth]{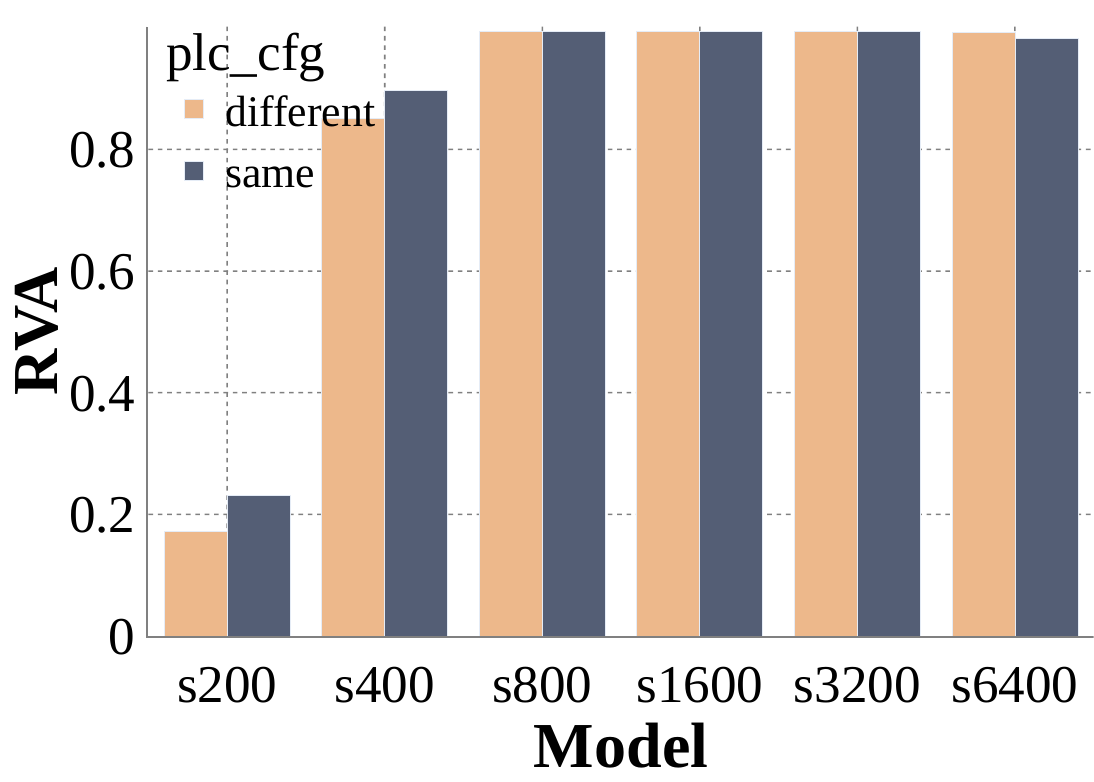}\label{subfig:protocol_emulation:s7comm:validator:emulation_dataset}}

\caption{Best \bca~and \rva~of the byt5-small model when using different dataset sizes to fine-tune for same and different configuration compared to the one used to fine-tune the model for generating the test dataset.}
\label{fig:protocol_emulation:emulation_dataset}
\end{figure}

Finally, we analyze the emulation datasets for different fine-tuning dataset sizes on both Modbus and S7comm protocols and demonstrate the results in Figure~\ref{subfig:protocol_emulation:exact:emulation_dataset} and Figure~\ref{subfig:protocol_emulation:s7comm:exact:emulation_dataset} for \bca~and Figure~\ref{subfig:protocol_emulation:validator:emulation_dataset} and Figure~\ref{subfig:protocol_emulation:s7comm:validator:emulation_dataset} for \rva. The most important outcome of the emulation is depicted in Figure~\ref{subfig:protocol_emulation:exact:emulation_dataset} and Figure~\ref{subfig:protocol_emulation:s7comm:exact:emulation_dataset}, where the \bca~is pretty low for different PLC configurations. In contrast, the \rva~is high since the validator incorporates logic that compares against the PLC configuration used when training the model. The performance across various configurations being unaffected by the size of the training dataset demonstrates that our results are independent of dataset size variety.

\begin{takeaway}
The model ``cloned'' the PLC with a specific configuration and behaves in the same manner and not in a random way in terms of responses.
\end{takeaway}

\subsubsection*{Varying supported function codes}
\begin{figure}[!t]
\vspace{-1.2em}
\subfloat[\bca]{\includegraphics[width=0.5\linewidth]{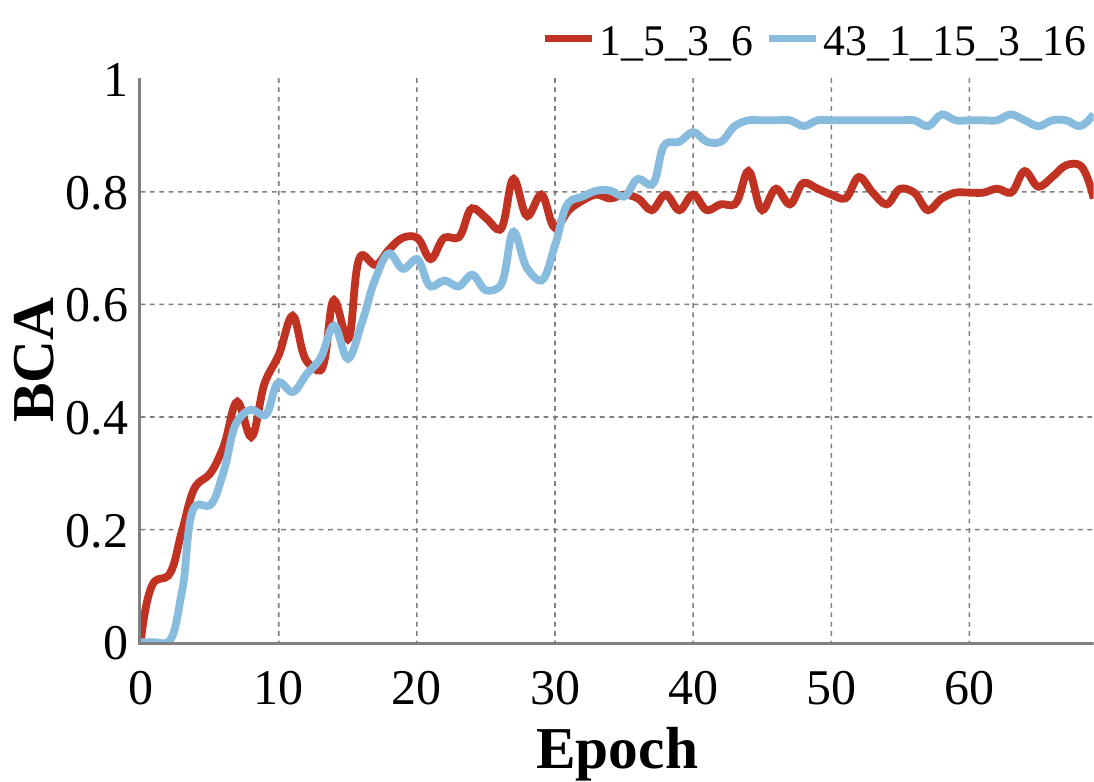}\label{subfig:diff-functions:exact}}
\subfloat[\rva]{\includegraphics[width=0.5\linewidth]{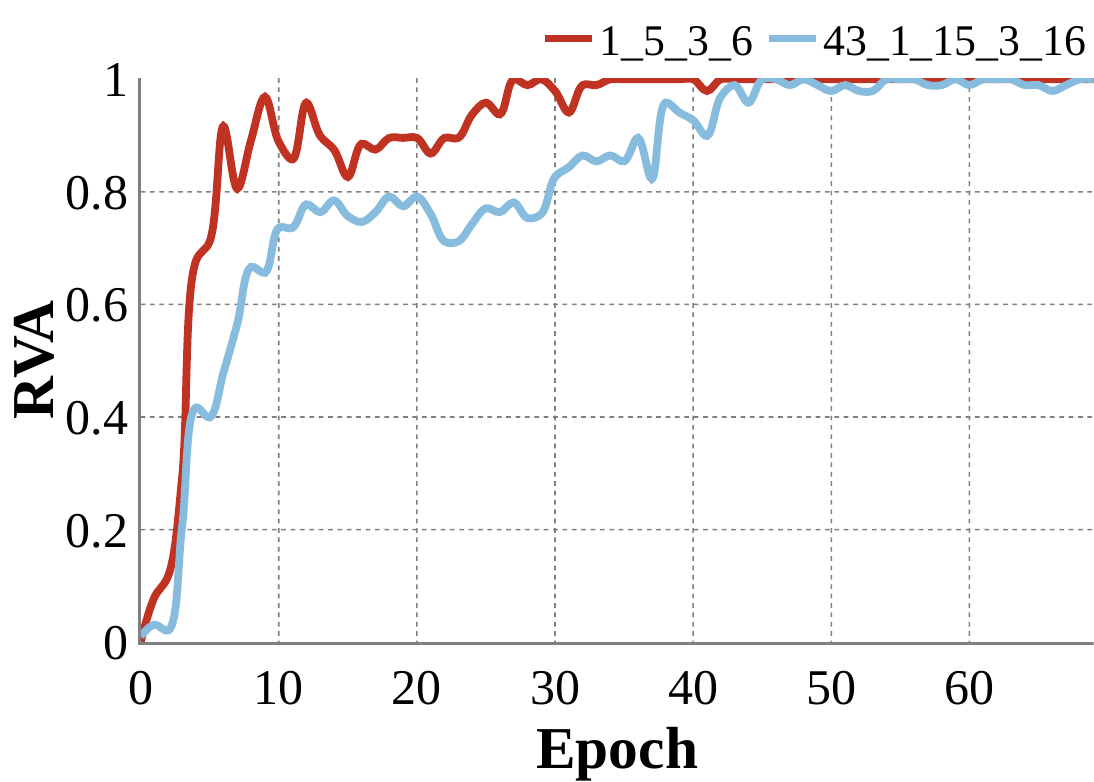}\label{subfig:diff-functions:validator}}

\caption{\bca~and~\rva~for different set of functions for Modbus protocol.}
\label{fig:diff-functions}
\end{figure}

Another experiment consists of two sets of various FCs, illustrated in Figure~\ref{fig:diff-functions} using red and cyan colors. The red set includes Read Discrete Coils (FC 1), Write Single Coil (FC 5), Read Holding Registers (FC 3), and Write Single Register (FC 6). The cyan set is composed of Device Identification (FC 43), Read Discrete Coils (FC 1), Write Multiple Coils (FC 15), Read Holding Registers (FC 3), and Write Multiple Registers (FC 16). For \bca, both sets yield similar final values, whereas for \rva, both achieve maximum values. Additionally, it is observed that convergence for the cyan set is slower, which is anticipated due to the set containing requests that involve multiple elements of digital or analog input(s)/output(s) (FC 15 and FC 16). This complexity contrasts with the red set, which comprises requests with a single element of digital or analog input(s)/output(s) and one additional function (FC 43).

\begin{takeaway}
\shortacronym~cloning is function independent and can achieve optimum performance in any set of functions.
\end{takeaway}

Overall, these results provide strong evidence related to Research Question (RQ1) in Section~\ref{subsection:research_questions}, proving that \shortacronym~effectively emulates the behavior of industrial network communication protocols with high realism.

\begin{takeaway}\label{takeaway:protocol}
The successful emulation of both Modbus and S7comm protocols by any ByT5 model size demonstrates \shortacronym's ability to generalize to different ICS protocols and PLC configurations using the proposed dataset generation method.
\end{takeaway}

%% file: sections/06-process_emulation.tex
In an ICS framework, control logic reflects the real-time interaction between devices and components involved in a physical process. Mathematical functions are the basic components of a physical process. Thus, exhibiting the LLM's ability to understand and replicate these functions is crucial for accurately mimicking physical processes.

\subsection{Exploration of ByT5 Approximating Control Functions}
\label{subsection:math_exploration}
Recent work is exploring the ability of LLMs to learn math, with varying degrees of success~\cite{ahn2024large}, \cite{ying2024internlm}. In this work, we explore ByT5's capability to approximate functions that typically appear in control logic, in the context of deceiving attackers. Thus, our aim is not to fully replicate math functions, rendering our scope much more limited than works focusing on LLMs learning math. Such efforts are orthogonal to our work, and can be incorporated if accurate math emulation is required. 

To explore ByT5 model's ability to approximate various mathematical functions, we utilized the Open Source Community for Automation Technology (OSCAT) library~\cite{oscat}, as it provides a wide range of functionalities that are used for industrial automation projects. This library was accessed using the CODESYS~\cite{codesys} platform running on a WAGO PFC200 PLC (shown in Figure~\ref{fig:plc_devices}). The mathematical functions included in this study were classified into discrete methods, involving conditional statements (signum function ``$sgn$"), and continuous methods, subdivided into exponential (exponential base 10 ``$expo10$" and hyperbolic cosine ``$cosh$"), and bounded (``$sigmoid$" and ``$cauchy$") categories.

When trying to ``clone'' the behavior of a PLC, it may not be possible to know apriori the mathematical function it uses. A practical approach would be to sample the PLC's responses across a range of uniformly spaced $x$-values to create a dataset. This dataset can then be used to train an LLM to approximate the PLC's function. However, uniform sampling may not effectively capture regions where the function changes more rapidly. By focusing on the regions where the function's slope varies significantly, we can optimize our sampling strategy for generating a dataset representing the function's characteristics more accurately. The optimized dataset helps estimate a probability density function (PDF), highlighting areas where denser sampling is required to capture the function's rapid changes accurately.

\begin{figure}[!t]
\vspace{-1em}
\subfloat[Sigmoid]{\includegraphics[width=0.5\linewidth]{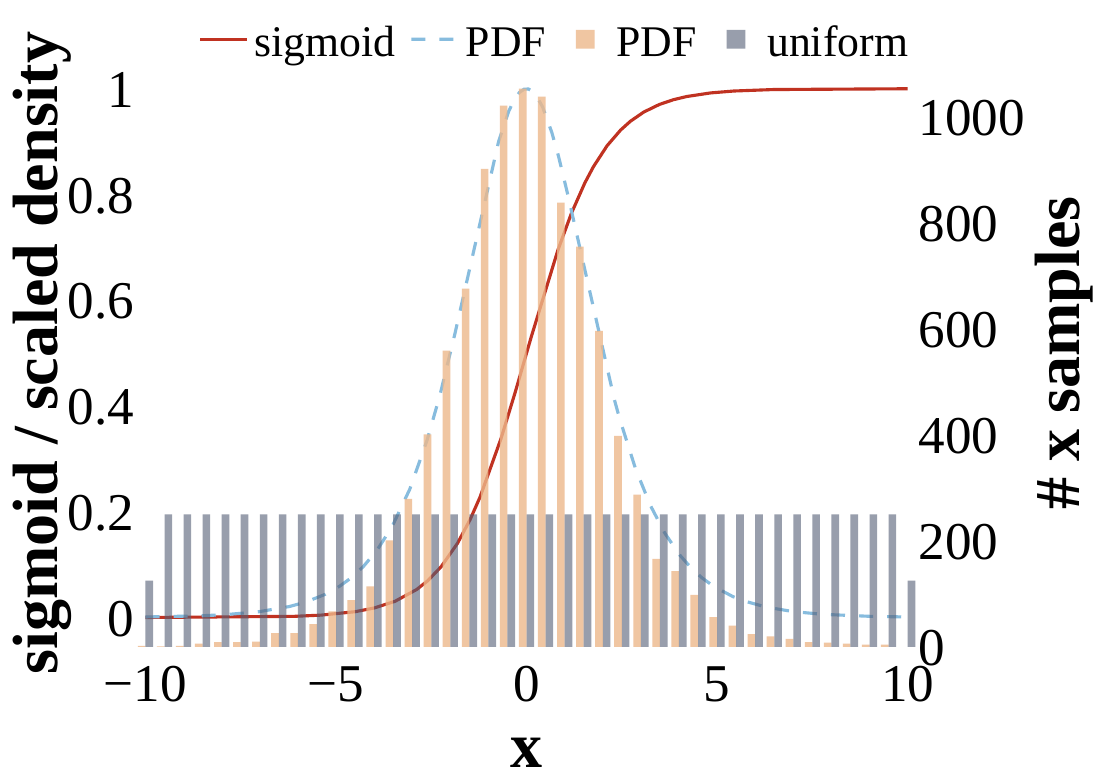}\label{subfig:math:sigmoid}}
\subfloat[Cauchy]{\includegraphics[width=0.5\linewidth]{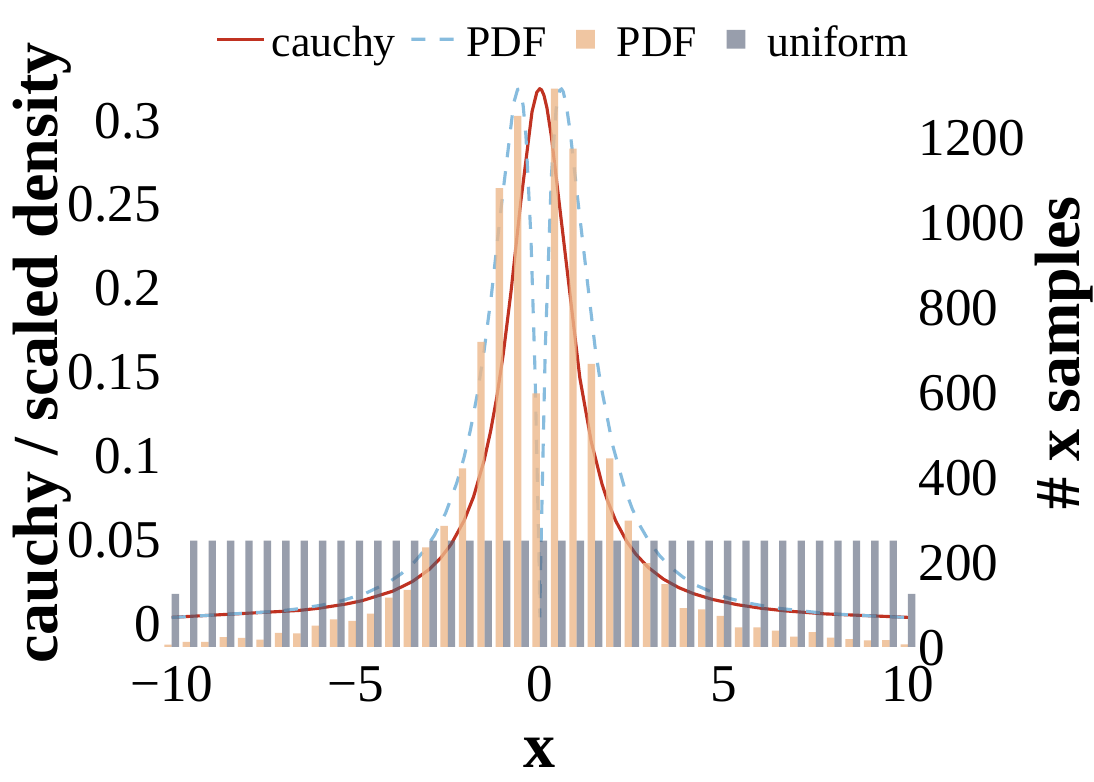}\label{subfig:math:cosh}}

\caption{Visual examples of the sampling method to sigmoid and cauchy functions. Plotted are the corresponding functions along with the derivative and $x$ axis bins.}
\label{fig:physical_process:math_functions}
\end{figure}

For example, flat regions of a function might be well represented by just a few points, whereas more complex curves could require many more, ideally an infinite number. This concept is illustrated with $sigmoid$ and $cauchy$ functions, as shown in Figure~\ref{fig:physical_process:math_functions}. In the figure, a histogram illustrates the distribution of sampled points along the $x$-axis, with yellow representing points sampled using the PDF mentioned above and gray indicating a discrete uniform distribution. We can observe that the PDF sampling is denser in areas where the functions change rapidly with $x$ and sparser where the function values are more stable. Algorithm~\ref{alg:math_functions} outlines the strategy for generating this weighted dataset. The main idea is to create a distribution of points on the $x$-axis that best represents the $y$-axis values of a mathematical function.

\begin{algorithm}[!t]
\caption{\small Algorithm to generate a weighted distribution of x-axis points for the math function explored.}
\label{alg:math_functions}
\begin{algorithmic}[1]
\footnotesize

\Require~~\\
$mix\_ratio$: parameter that dictates the weighting between the uniform and derivative-based probability density in the mixed PDF. \\
$power$: value between (0, 1] that softens high values of the $df$. \\
$df$: derivative of the math function studies, returns a set of $x,\, y$. \\
$ct$: function that returns the cumulative integral of $y$ along $x$. \\
$inverse\_cdf$: function that returns values of $x$ corresponding to the cumulative density values using interpolator.
\\
\Function{x\_axis\_values\_distributed}{$x\_values$}
    \State $samples \gets \Call{len}{x\_values}$                                        \label{alg:x_axis:samples}
    \State $dv \gets$ \Call{df}{$x\_values,\,samples$}                                  \label{alg:x_axis:dv}
    \State $dv \gets \Call{power}{dv,\,power}$ 
    \State $diff \gets dv[0] - dv[1]$                                                   \label{alg:x_axis:diff}
    \State $pdf \gets \frac{dv}{\Call{sum}{dv \cdot diff}}$                             \label{alg:x_axis:pdf}
    \State $uni \gets \Call{ones}{samples} / samples$                                   \label{alg:x_axis:ones}
    \State $adj\_pdf \gets (1 - mix\_ratio) \cdot dv + mix\_ratio \cdot uni$           \label{alg:x_axis:adj_pdf}
    \State $adj\_pdf \gets \frac{adj\_pdf}{\Call{sum}{adj\_pdf \cdot diff}}$            \label{alg:x_axis:adj_pdf_norm}
    \State $cdf \gets \Call{ct}{adj\_pdf,\, x\_values}$                                 \label{alg:x_axis:cdf_ct}
    \State $cdf \gets cdf / cdf[-1]$                                                    \label{alg:x_axis:cdf_norm}
    \State $uni \gets \Call{random}{samples}$                                           \label{alg:x_axis:inverse_cdf_uni}
    \State $inv\_cdf \gets \Call{interp\_1d}{cdf,\, x\_values,\, 'linear'}$                     \label{alg:x_axis:inverse_cdf_inter}
    \State $x \gets inv\_cdf(uni)$                                                      \label{alg:x_axis:inverse_cdf_values}
    \State $x \gets \Call{sort}{x}$                                                     \label{alg:x_axis:sort}
    \State \Return $x$
\EndFunction

\end{algorithmic}
\end{algorithm}

This algorithm requires uniformly distributed points as a base for sampling along the $x$-axis to facilitate the application of the Inverse Transform Sampling (ITS) method~\cite{sheldon2018first}. It starts by calculating the derivative of the target function which is numerically estimated based on the responses of the PLC for an unknown function [line:~\ref{alg:x_axis:dv}], which is then raised to a selected \textit{power} between zero and one to soften its values, thus reducing the amount of points incorporated in regions of rapid change. The derivative is normalized [line:~\ref{alg:x_axis:pdf}] using the difference between two uniformly distributed points [line:~\ref{alg:x_axis:diff}], resulting in a PDF. To avoid zeros when the function flattens and ensure a comprehensive exploration of the input space, a mixed PDF is constructed by combining the uniformly distributed PDF with the derivative-based PDF, controlled by the \textit{mix\_ratio} [line:~\ref{alg:x_axis:adj_pdf}]. This ratio adjusts the contribution of each PDF to the mix, focusing on critical regions while covering the entire range. The resulting adjusted PDF is normalized [line:~\ref{alg:x_axis:adj_pdf_norm}] by dividing it by the sum of the product of the mixed PDF and the interval between successive $x$-values [line:~\ref{alg:x_axis:diff}], ensuring the area under the curve equals one and forms a valid probability distribution.
To derive the adjusted $x$-axis data points from the adjusted PDF, we first calculate the cumulative distribution function (CDF) by integrating the PDF across uniformly distributed $x$-values using the trapezoidal rule [line:~\ref{alg:x_axis:cdf_ct}]~\cite{burden2016numerical}. This step forms the basis for generating the inverse CDF through linear interpolation [line:~\ref{alg:x_axis:inverse_cdf_inter}], using the initial distribution of $x$-values. The inverse CDF then interacts with a uniformly distributed set of values to produce the adjusted $x$-axis data points, which are subsequently sorted [line:~\ref{alg:x_axis:sort}] for the sake of clarity.

\begin{figure}[!t]
\vspace{-1em}
\subfloat[Sigmoid]{\includegraphics[width=0.5\linewidth]{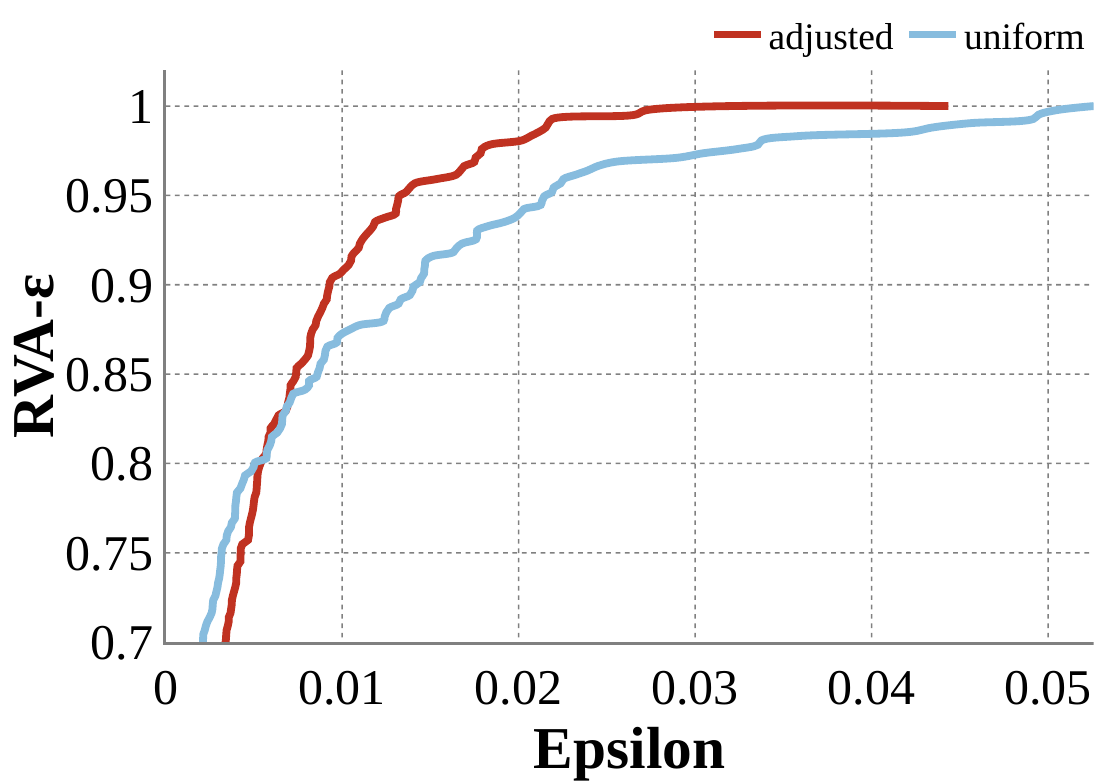}\label{subfig:math:sigmoid:rvae}}
\subfloat[Cauchy]{\includegraphics[width=0.5\linewidth]{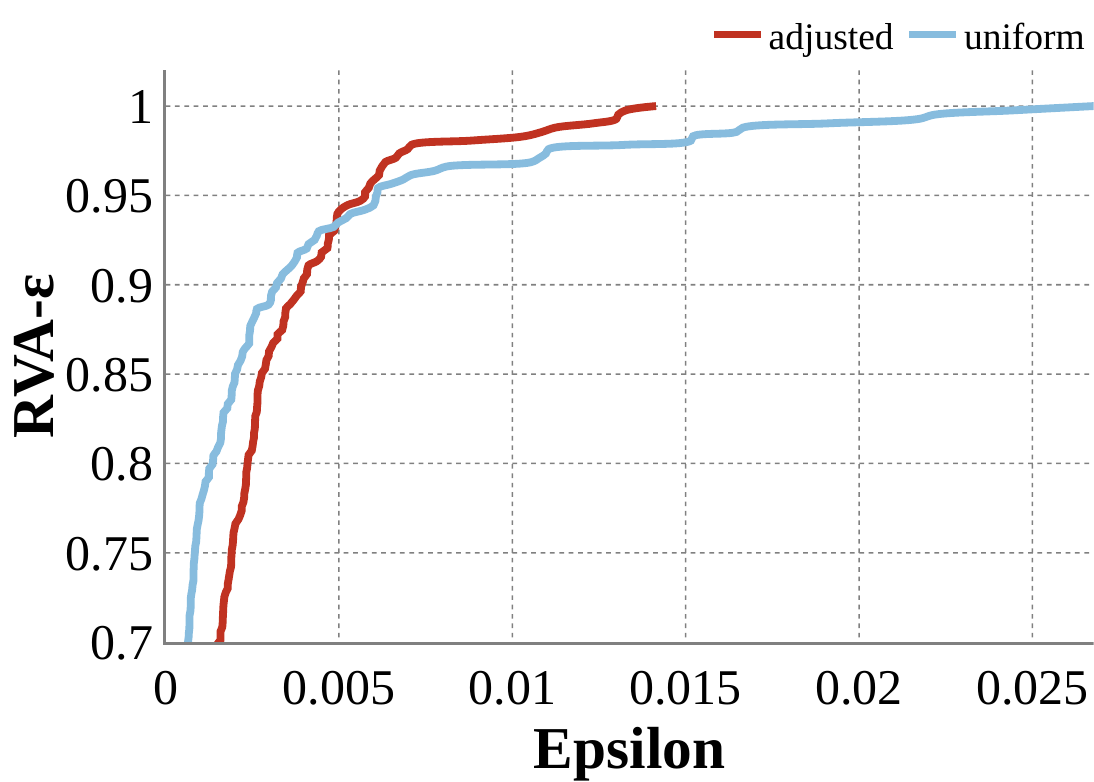}\label{subfig:math:cauchy:rvae}}
\hfill
\subfloat[Exponential base 10]{\includegraphics[width=0.5\linewidth]{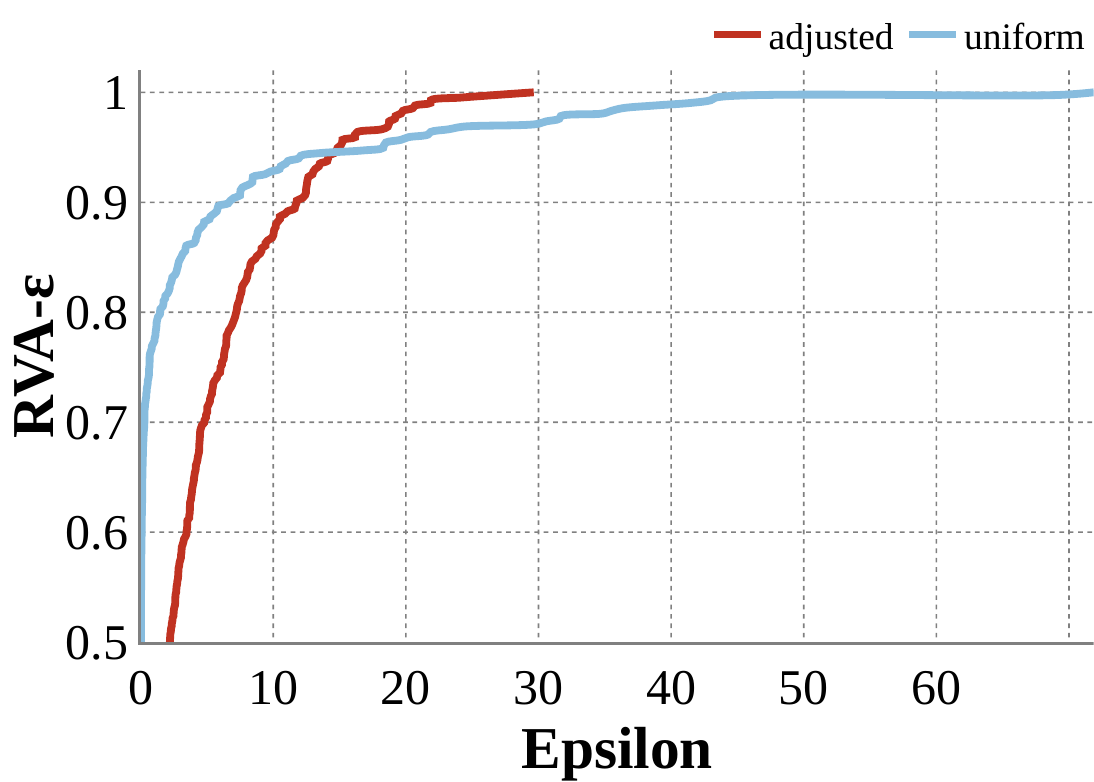}\label{subfig:math:expo10:rvae}}
\subfloat[Hyperbolic Cosine]{\includegraphics[width=0.5\linewidth]{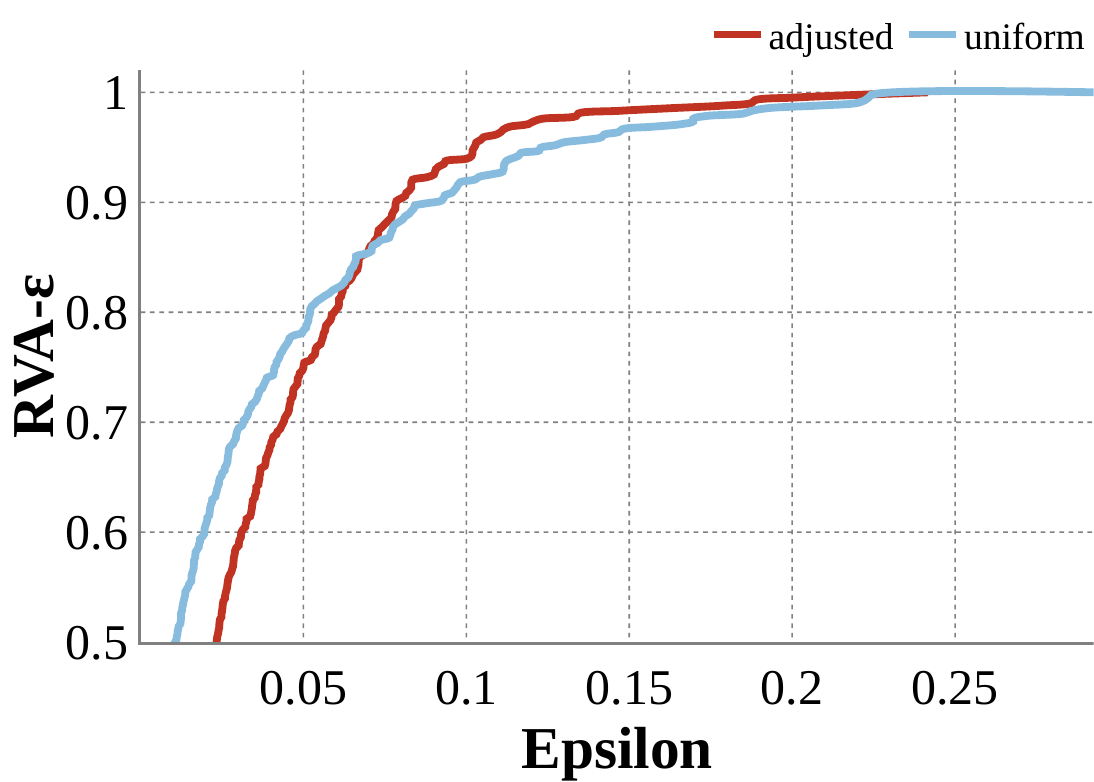}\label{subfig:math:cosh:rvae}}

\caption{Context, \rvae~for math functions. Red shows the metric using Algorithm~\ref{alg:math_functions} distribution; cyan shows the result with the uniform discrete distribution.}
\label{fig:math:functions:rva-e}
\end{figure}

The output of Algorithm~\ref{alg:math_functions} is used in an iterative loop to probe the PLC with the sampled $x$-axis values to generate necessary packets for model fine-tuning. This loop is much simpler than the one used for protocol emulation in Section~\ref{section:protocol_emulation}. It consists of a write and then a read on the analog input/output for each sample collected by the algorithm and controlled by the range that communication protocols (e.g., Modbus, S7comm, etc.) can support. The amount of collected data follows the same method discussed in Section~\ref{section:protocol_emulation}.

\subsubsection*{Results}
Figure \ref{fig:math:functions:rva-e} depicts the results obtained from experimenting with the capability of the LLM model to emulate math functions. The $x$-axis and $y$-axis represent the epsilon value and the cumulative \rvae, respectively. Starting with Figure~\ref{subfig:math:sigmoid:rvae} for $sigmoid$ and Figure~\ref{subfig:math:cauchy:rvae} for $cauchy$, Algorithm~\ref{alg:math_functions} manages to create an advantage over the uniform sampling method by increasing \rvae~metric faster and providing full accuracy in smaller epsilon. This effect is seen to be bigger in exponential functions, like $expo10$ in Figure~\ref{subfig:math:expo10:rvae} and $cosh$ in Figure~\ref{subfig:math:cosh:rvae}. For these functions, the standard deviation is significantly smaller for the adjusted distribution, 5.1 for \textit{expo10} and 0.036 for \textit{cosh}, while it is 8.23 and 0.045 respectively when using a uniform distribution. This is depicted by the fact that \rvae~curve reaches its maximum sooner. On the other hand, the $sgn$ outputs -1, 0, or 1 depending on whether the input is negative, zero, or positive respectively. Due to the simplicity of the function's conditional nature, the derivative used to calculate PDF will always be zero. This allows \shortacronym~to quickly capture the pattern and emulate the function accordingly with \bca~and \rva~of $100\%$.

\begin{takeaway}\label{takeaway:math}
A weighted approach outperforms uniform sampling for LLM-based math function approximation. \shortacronym~excels with simple and bounded functions, while unbounded ones remain challenging—though acceptable given the target honeypot application since PLC variables are inherently constrained.
\end{takeaway}

\subsection{ByT5 Emulation of Physical Processes}
Building on the previous analysis, we extend our exploration to evaluate ByT5's capabilities in emulating various control programs. We apply this evaluation to five industrial processes from ICSPatch~\cite{rajput2023icspatch}: \textit{Aircraft Flight Control}, \textit{Anaerobic Digestion Reactor}, \textit{Chemical Plant}, \textit{Desalination Plant}, and \textit{Smart Grid}. These systems were accessed through CODESYS on a WAGO PFC200 PLC using the Modbus communication protocol.

The dataset generation for control logic processes follows Algorithm~\ref{alg:boundaries_client} by using read and write commands towards the PLC to capture the inherent functionality of the physical process. The values for the write commands are derived from evenly distributed values within a defined interval. This was initially random but later adjusted incrementally to ensure consistent model performance across different intervals.

\begin{figure}[!t]
\vspace{-1em}
\subfloat[\bca~on Modbus protocol]{\includegraphics[width=0.5\linewidth]{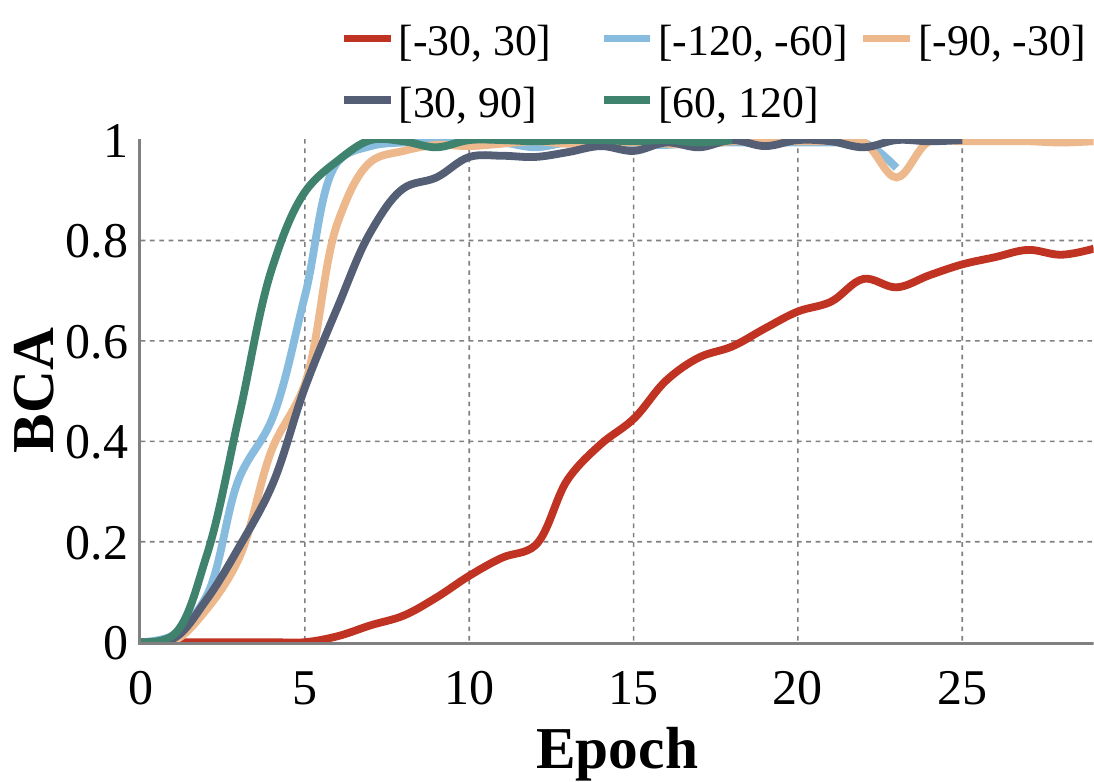}\label{subfig:ablation:aircraft:bca}}
\subfloat[\rva~on Modbus protocol]{\includegraphics[width=0.5\linewidth]{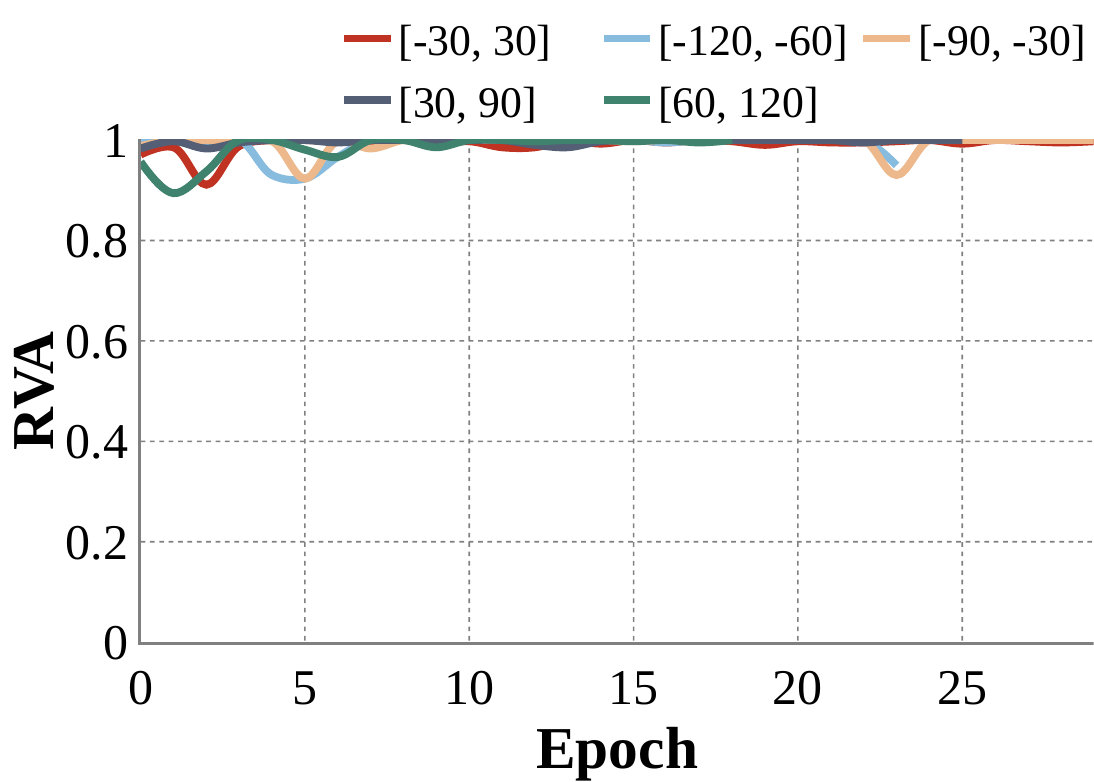}\label{subfig:ablation:aircraft:pva}}

\caption{\bca~and \rva~per epoch of byt5-small model for different probe value ranges for the ICSPatch anaerobic physical process using contextual prompt.}
\label{fig:ablation:aircraft:bca-pva}
\end{figure}

\begin{figure}[!t]
\vspace{-1em}
\subfloat[\bca~on ICSPatch processes]{\includegraphics[width=0.5\linewidth]{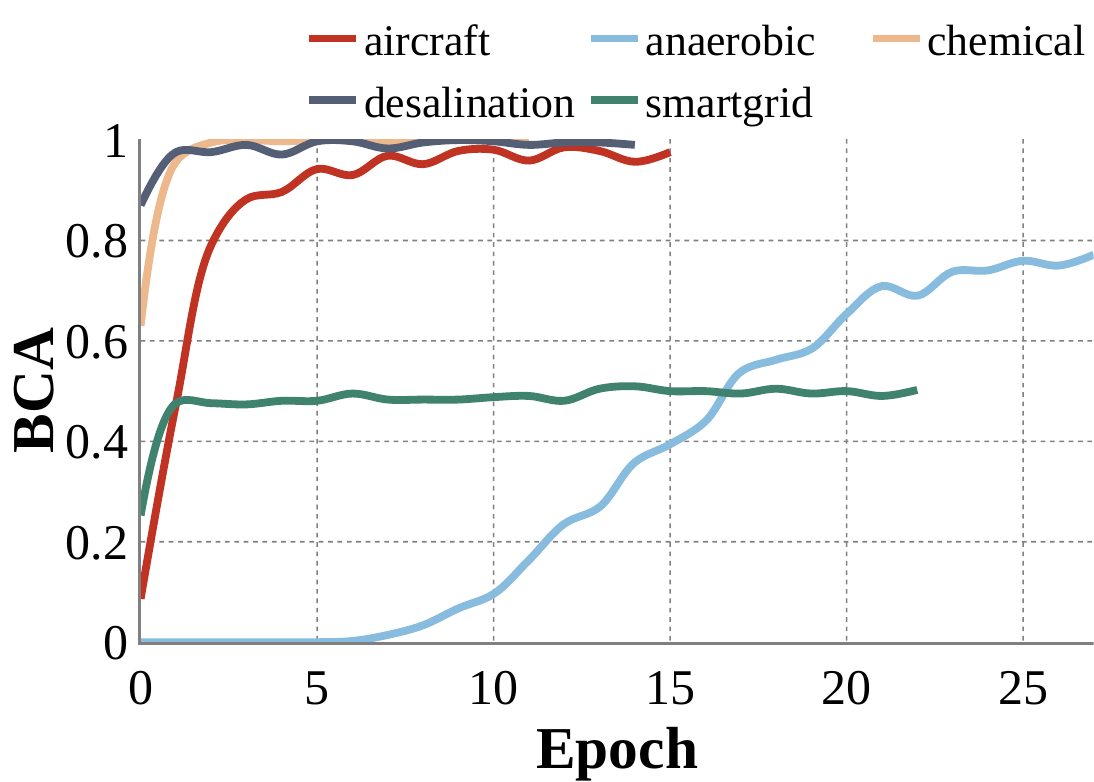}\label{subfig:icspatch:bca}}
\subfloat[\rva~on ICSPatch processes]{\includegraphics[width=0.5\linewidth]{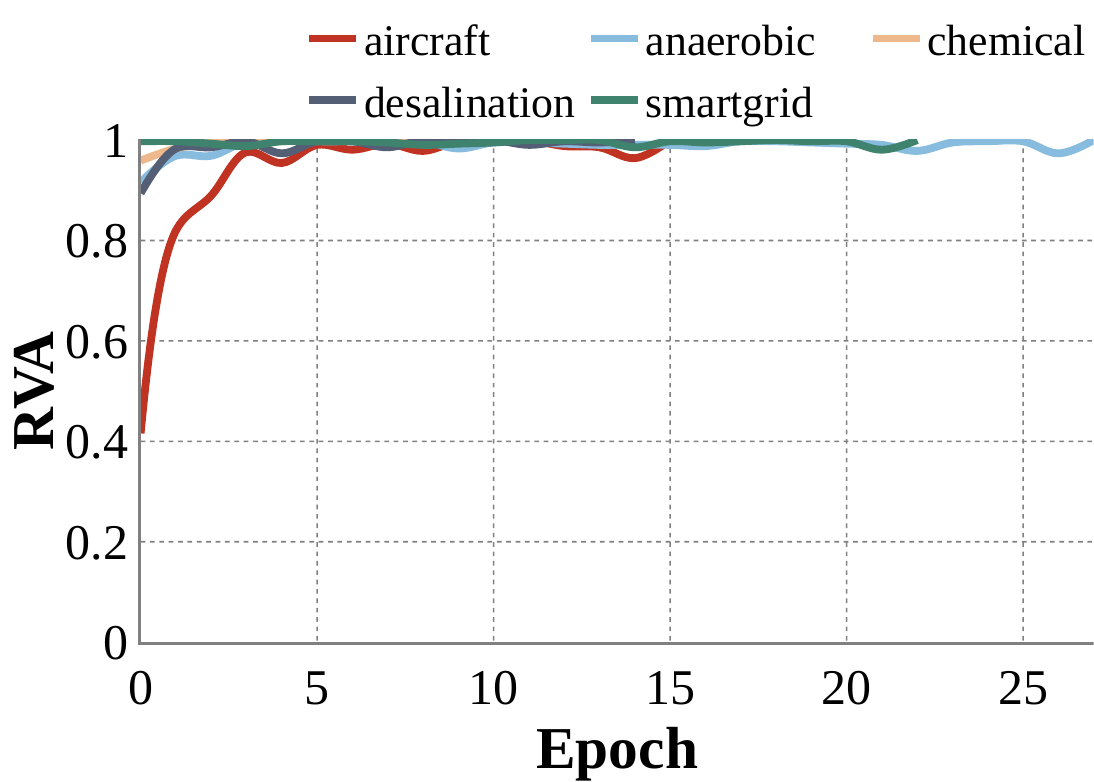}\label{subfig:icspatch:pva}}\hfill

\caption{\bca~and \rva~per epoch of byt5-small model emulating ICSPatch processes using contextual prompt.}
\label{fig:icspatch:bca-pva}
\end{figure}

\subsubsection*{Results}
Figure~\ref{fig:ablation:aircraft:bca-pva} depicts~\shortacronym's ability to effectively emulate a physical process regardless of the value ranges. Depending on the value range selected, the physical process has different behavior and this is depicted in the range [-30, 30]. The fluctuation of the returned value is more rapid and thus the \bca~is not reaching 100\% as the other ranges do. Furthermore, Figure~\ref{fig:icspatch:bca-pva} represents the capability of the byt5-small model to emulate control logic processes from the ICSPatch~\cite{rajput2023icspatch} case study. The results of~\bca~and ~\rva~shown in Figure~\ref{subfig:icspatch:bca} and Figure~\ref{subfig:icspatch:pva}, respectively, prove the ability of \shortacronym~to capture the states of the variables across multiple processes. However, it can be noted that the emulation for the anaerobic (value range [-30,30]) and smart grid processes includes some complex functionality and large input and output numbers, which explains the reduction in performance as depicted in Figure~\ref{subfig:icspatch:bca}. This requires additional ablation studies to improve the dataset generation to capture the variables' states accurately.

\begin{takeaway}\label{takeaway:process1}
Physical processes can achieve high~\bca~and~\rva~similarly to protocol emulation in Section~\ref{section:protocol_emulation} proving that context prompt does not impact the performance of the model.
\end{takeaway}

\begin{takeaway}\label{takeaway:process}
Consistent with the math function exploration, \shortacronym~can capture effectively processes with limited input/output range. Capturing processes with more complex functionality require deeper knowledge of the process to properly guide the dataset generation strategy.
\end{takeaway}

%% file: sections/07-case_study.tex
To understand the efficiency of \shortacronym~in emulating real-world industrial processes, we experiment with a Multi-Stage Flash (MSF) desalination plant testbed~\cite{rajput2019process}. This process consists of various components such as flow rate, temperature and pressure sensors, valves, and PID controllers. In the setup, MATLAB Simulink simulates the desalination process, while the connected PLCs, in Hardware-In-The-Loop (HITL) setup, manage part of the physical process by receiving the Initial Brine Temperature (TB0) and the Distillate Product Flow Rate (Wd) as inputs to regulate the Steam Flow Rate (Ws) using a cascaded PID controller setup~\cite{ali2002understanding}.

In the default setup, the PID set-point value is set to 93, directly affecting the steam flow rate and producing a steady-state value. During data collection, the set-point value, the timestamp, and the steam flow rate value were integrated into the dataset's context and used as input to the \shortacronym~fine-tuning process. For testing, minimal variations were randomly added to timestamps to avoid overlapping with the training set, and random set-point values were used at different timestamps to monitor \shortacronym~emulation results. 

Through experimentation, it was noted that set-point values above 85, closer to 93, showed minimal to no fluctuation in Ws. Therefore, we decided not to experiment with larger set-point values. Furthermore, we observed that with abnormally low set-point values, below 70, \shortacronym~starts to fail. The high number of fluctuating points affected by the low set-point values make it challenging for \shortacronym~to capture the disturbance fluctuations accurately and thus resulted in the model averaging the data points and losing the Ws steady-state value (refer to Figure~\ref{fig:testbed:time-sp65} in Appendix~\ref{appendix:testbed}). Thus, a trade-off must be made between fine-tuning \shortacronym~for all possible set-point values and maintaining the Ws steady-state value. Since the ultimate objective is to emulate the physical process and to accurately reflect the values, four set-point values with intervals of 5 points (85, 80, 75, and 70) were chosen for conciseness. 

\begin{figure}[!t]
\vspace{-1em}
\subfloat[dataset size: 1600]{\includegraphics[width=0.5\linewidth]{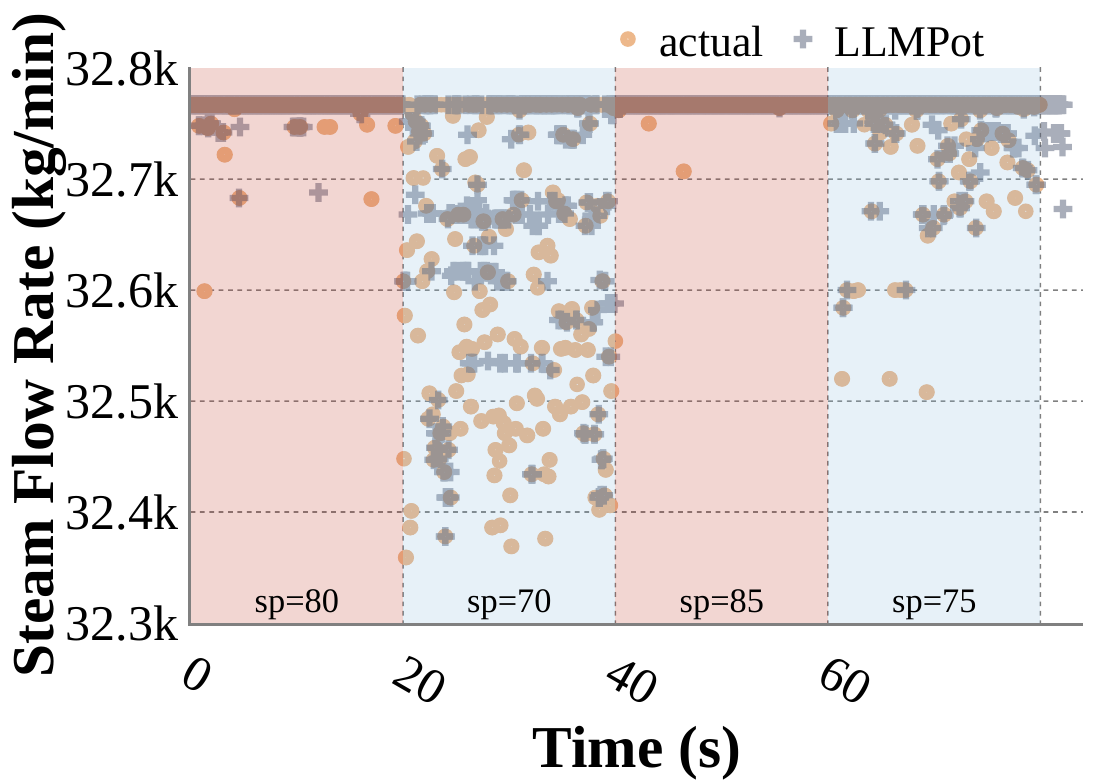}\label{subfig:testbed:sp00:s1600}}
\hfill
\subfloat[dataset size: 3200]{\includegraphics[width=0.5\linewidth]{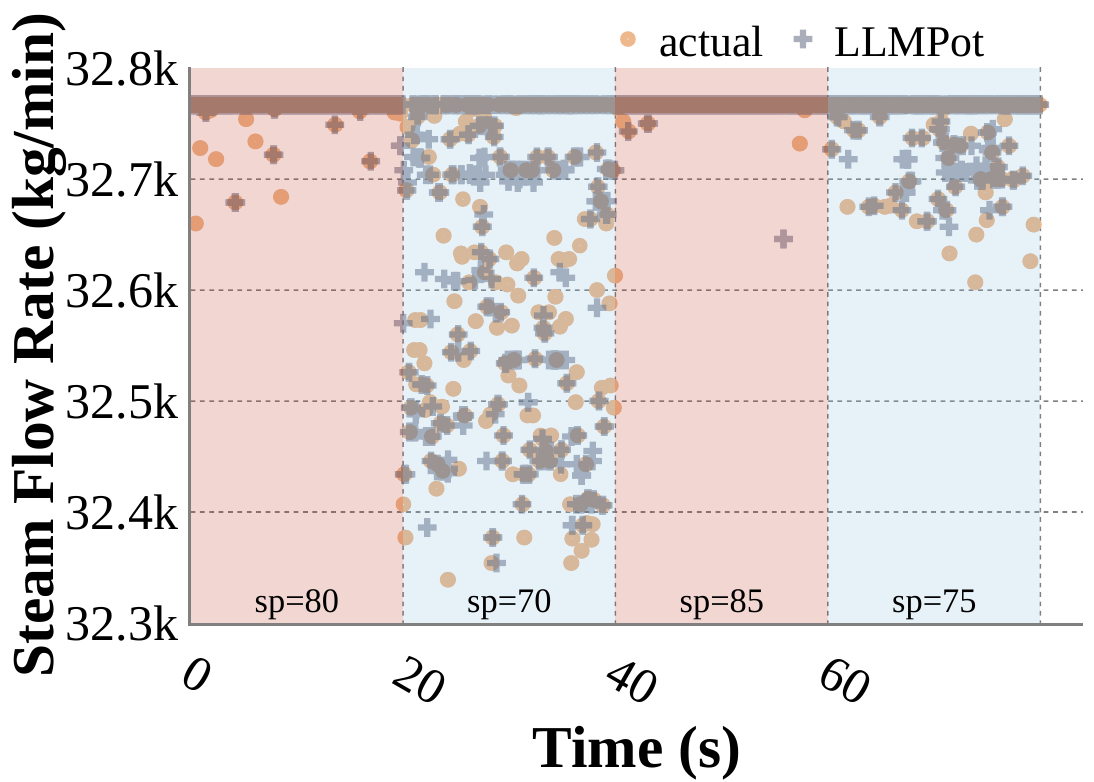}\label{subfig:testbed:sp00:s3200}}
\caption{Steam Flow Rate (Ws) fluctuations over time during set-point changes across trials.
}
\label{fig:testbed:time-sp}
\end{figure}

Figure~\ref{fig:testbed:time-sp} shows the distribution of the data points against the time axis taken from the testbed and the emulated results overlapped, where at each time interval a random set-point value was sent to \shortacronym. Figure~\ref{subfig:testbed:sp00:s1600} and Figure~\ref{subfig:testbed:sp00:s3200} show results with two different dataset sizes, 1600 and 3200 used for training~\shortacronym, respectively. It is worth noting that with a dataset size below 1600, \shortacronym~fails to predict points beyond the Ws steady-state value, as it does not include enough fluctuations for \shortacronym~to be trained properly. Conversely, with a larger dataset size (3200 samples), more disturbance points were included, providing a greater advantage in fine-tuning \shortacronym. Furthermore, it can be seen that the majority of the data points from the testbed and \shortacronym~are clustered around the Ws steady-state value. Results depict that for different set-points, \shortacronym~is capable of learning the range and general distribution and emulating the disturbance fluctuation of a real testbed within a normal operational range.\footnote{Additional results on different static set-point values appear in Figure~\ref{fig:appendix:testbed:time} in Appendix~\ref{appendix:testbed}.}

To statistically prove this, we measured the similarity and calculated the p-value for both real and emulated distributions using the Kolmogorov-Smirnov test (KS-test)~\cite{daniel1990applied}. The experiment was repeated 10 times and results show p-values above 0.05 indicating no significant difference between both distributions (refer to Table~\ref{table:statistics} in Appendix~\ref{appendix:testbed}).

\begin{takeaway}\label{takeaway:testbed}
\shortacronym~can provide suitable steady-state responses over time, effectively capturing the effect of disturbances in a real testbed and providing the output fluctuations an attacker could be observing.  Nonetheless, Section~\ref{subsection:math_exploration} also proves that the $x$-axis can potentially be the time dimension in a physical process.
\end{takeaway}

%% file: sections/08-related_work.tex
Honeypots are typically categorized as either \textit{low-interactive} or \textit{high-interactive}, based on their realism and interaction levels.

\textbf{Low-interactive honeypots} This type simulates essential limited services and are cost-effective and relatively easy to deploy. CryPLH~\cite{buza2014cryplh} explicitly emulates the S7comm protocol using a manually implemented Python script by querying a specific PLC with a particular configuration and integrating these values into the script. Unfortunately, the source code for this project is not available for further investigation. Furthermore, SHaPe~\cite{koltys2015shape} focuses on emulating the IEC 61850 MMS protocol using a library libiec61850 and a framework called Dionaea, however, this project has been discontinued. Gaspot~\cite{wilhoit2015gaspot} implements Guardian AST protocol using Python scripts that reply to specific requests for a use-case scenario (gas station) implemented for the project. Moreover, Conpot~\cite{jicha2016scada} supports emulating multiple protocols using Python scripting. For Modbus, the implementation supports FCs 1,2,3,4,5,6,15,16,17,43 by simulating the PLC's memory so their replies always comply with how a PLC device would work. This work supports XML-based configuration but still lacks the ability to modify the behavior of the honeypot for specific scenarios, that would need additional Python coding. Overall, the simplicity and limited functionality of the aforementioned \emph{low-interactive honeypots lead the attackers to easily identify them as decoys} \cite{lopez2020honeyplc}.

\textbf{High-interactive honeypots} On the other hand, this type provides a more complex and authentic environment that simulates real operating systems and their services, making them ideal for detailed data collection on attackers' methodologies. Key examples of high-interactive honeypots include DiPot~\cite{cao2018dipot}, a distributed system that uses Conpot~\cite{jicha2016scada} to simulate industrial protocols. S7commTrace~\cite{xiao2018s7commtrace} claims superiority over Conpot by extending the S7comm protocol with additional FCs. However, the source code for both DiPot and S7commTrace is not available. Furthermore, HoneyPLC~\cite{lopez2020honeyplc} enhances functionality by simulating the storage of ladder logic on PLCs and supports out-of-the-box PLCs such as Siemens S7-300, Allen-Bradley MicroLogix 1100, ABB PM554-TP-ETH PLCs, etc. Their implementation utilizes the Snap7 library which was modified (following a combination of design patterns~\cite{martin2000design} like proxy or delegator could be a more appropriate way instead of modifying it) and recompiled to suit the updated functionality, making the honeypot very difficult to configure for different PLCs with different analog and digital inputs/outputs. NeuralPot~\cite{siniosoglou2020neuralpot} employs a Generative Adversarial Network~\cite{goodfellow2014generative} and Auto-Encoders~\cite{hinton2006reducing} to learn the behavior of the ICS device and dynamically generate Modbus traffic. They also utilize Conpot for their protocol emulation, however no source code is available. Overall, the aforementioned high-interactive honeypots \emph{lack the essential aspect of ensuring high fidelity by emulating a physical process.} 

\begin{table}[!t]
\centering
\caption{ICS honeypot comparison.}
\label{table:honeypots}
  \resizebox{\columnwidth}{!}{%
  \begin{tabular}{ | p{0.15cm} | >{\centering}p{2.35cm} | c | P{0.7cm} | P{0.2cm} | P{0.7cm} | c | p{0.3cm} | c | P{0.2cm} |}
    \hline
     \multirow{2}{*}{\thead{\rotatebox[origin=c]{90}{\textbf{Interactivity}}}} &
     \multirow{2}{*}[-2em]{\thead{\textbf{Honeypot}}} &
     \multirow{2}{*}[-2em]{\textbf{\thead{Protocol\\Emulation Base}}} &
     \multicolumn{3}{c|}{\thead{\textbf{Emulation}}} &
     \multicolumn{3}{c|}{\thead{\textbf{Extensibility}}} & 
     \multirow{2}{*}{\thead{\rotatebox[origin=c]{90}{\textbf{Open Source}}}} \\
     \cline{4-9}
      & & & 
     \thead{\rotatebox[origin=c]{90}{\textbf{Protocol}}} &
     \multirow{1}{*}[2em]{\thead{\rotatebox[origin=c]{90}{\textbf{Ctrl Logic}}}} &
     \thead{\rotatebox[origin=c]{90}{\textbf{Accuracy}}} &
     \thead{\rotatebox[origin=c]{90}{\textbf{PLC}}} &
     \thead{\rotatebox[origin=c]{90}{\textbf{Protocol}}} &
     \thead{\rotatebox[origin=c]{90}{\textbf{Ctrl Logic}}} & \\
    \hline
    \multirow{4}{*}{\rotatebox[origin=c]{90}{\textit{low}}}
    & CryPLH~\cite{buza2014cryplh}                 & CryPLH & Single & - & 100 & \wc & \wc & \wc & \cmark \\    
    & SHaPe~\cite{koltys2015shape}                 & SHaPe  & Single & - & 100 &\lc & \wc & \wc & \xmark \\
    & Gaspot~\cite{wilhoit2015gaspot}              & Gaspot & Single & 1 & 100 &\wc & \wc & \wc & \cmark \\
    & Conpot~\cite{jicha2016scada}                 & Conpot & Multi & - & 100 &\lc & \lc & \wc & \cmark \\
    \cline{1-10}
    \multirow{8}{*}{\rotatebox[origin=c]{90}{\textit{high}}}
    & DiPot~\cite{cao2018dipot}                    & Conpot~\cite{jicha2016scada}  & Multi & - & 100 &\lc & \wc & \wc & \xmark \\
    & S7commTrace~\cite{xiao2018s7commtrace}       & Conpot~\cite{jicha2016scada}  & Single & - & 100 &\lc & \wc & \wc & \xmark \\
    & HoneyPLC~\cite{lopez2020honeyplc}            & snap7~\cite{snap7} & Single & - & 100 &\bc & \wc & \wc & \cmark \\
    & NeuralPot~\cite{siniosoglou2020neuralpot}    & Conpot~\cite{jicha2016scada} & Single & - & 100 &\lc & \wc & \wc & \xmark \\
    & Antonioli et al.~\cite{antonioli2016towards} & MiniCPS~\cite{antonioli2015minicps} & Single & 1 & 100 &\lc & \wc & \wc & \cmark \\
    & HoneyPhy~\cite{litchfield2016rethinking}     & HoneyPhy~\cite{litchfield2016rethinking} & Single & 1 & 100 &\lc & \wc & \wc & \xmark \\
    & ICSpot~\cite{conti2022icspot}                & HoneyPLC~\cite{lopez2020honeyplc} & Multi & 1 & 100 &\bc & \wc & \wc & \cmark \\
    \cline{2-10}
    & \textbf{\shortacronym}                       & \textbf{\shortacronym} & \textbf{Multi} & \textbf{6+} & \textbf{99.30} &\bc & \lc & \lc & \cmark \\
    \hline
  \end{tabular}
  }
  \vspace{0.1cm}
  \par \bc:~Automated (modification of existing configuration),\\\lc:~Semi-automated (guided manual effort needed),\\\wc:~Not Extensible,  - (dash):~Not supported,\\Accuracy: Packet Validity Percentage (\rva)\\
\end{table}

Antonioli et al.~\cite{antonioli2016towards} addressed this by designing a virtual high-interactive honeypot that mimics a water treatment process. However, developing a testbed that accurately simulates real-world conditions or integrates various technologies is challenging and highly complex. On top of that, building testbeds is costly and requires regular maintenance and technical support to keep the testbed operational. HoneyPhy~\cite{litchfield2016rethinking} focuses on creating realistic cyber-physical system (CPS) simulations using the Distributed Network Protocol (DNP3) protocol~\cite{dnp3spec}. However, this work relies on constructing a white box model, which requires detailed knowledge of the ICS device, also no source code is available. ICSPot~\cite{conti2022icspot} is built on top of HoneyPLC~\cite{lopez2020honeyplc}, enhancing the emulation of network communications and physical-layer interactions using Industrial Hacking Simulator~\cite{lannister2022ihs}. It utilizes a simplified simulation of the water treatment process provided by MiniCPS~\cite{antonioli2015minicps}. The authors in~\cite{antonioli2015minicps} use the PyModbus Python library to emulate Modbus protocol by implementing the server side, customized with if-else statements. On top of that, they implement two distinct physical processes, however, the effort needed to implement these is proportional to the complexity of the processes itself. Overall, the aforementioned related works \emph{support only a single predetermined physical process.}

Table~\ref{table:honeypots} illustrates a qualitative comparison of the aforementioned state-of-the-art honeypots with our proposed \shortacronym. Interactivity column is an abstractly defined term in the literature and mostly adheres to the amount of time and data exchanged between a client/attacker and the honeypot. Each honeypot usually depends on a base software that performs the needed initialization and communications to emulate a custom PLC. The protocol emulation column is labeled as either ``Single'' or ``Multi'' based on whether the honeypot emulates one or more protocols. The control logic emulation column defines how many physical processes are emulated if any. Accuracy column reports the validity of the generated packet by the honeypot using the~\rva~metric.~\shortacronym~demonstrates highly comparable accuracy to the existing state-of-the-art honeypots. The 99.30\% is derived from Table~\ref{table:protocol_emulation:model_sizes} where the byt5-small model was trained using the 1600 samples dataset. Finally, the extensibility section qualitatively describes the ability of the honeypot to expand to additional PLC devices, protocols, and control logic. The PLC ``Extensibility'' term, defined in~\cite{lopez2020honeyplc}, allows effective PLC \emph{hardware} emulation of different models and vendors.

For both protocol and control logic extensibility, \shortacronym~manages to excel compared to other honeypots. It provides the user the ability to extend the functionality by following the methodology described in Algorithm~\ref{alg:boundaries_client} for protocols, and by probing physical processes as described in Section~\ref{section:process_emulation} for the overall goal of replicating the user chosen configuration for the honeypot.

\begin{takeaway}
    \shortacronym~becomes more dynamic and adaptable to various PLC configurations, moving away from the traditional reliance on static and manual scripting specific to PLC setups. Two distinct protocols and more than six physical processes were emulated as a case study.
\end{takeaway}

%% file: sections/09-honeypot_setup.tex
\subsection{Honeypot Setup}
\shortacronym~uses Nmap~\cite{orebaugh2011nmap} and wget~\cite{wget} to automatically gather essential parameters such as MAC addresses and operating systems from the target PLC device. We emulate the honeypot's TCP/IP stack using Honeyd~\cite{provos2003honeyd}, configuring it with the appropriate Nmap fingerprint and integrating real services via Docker containers~\cite{docker}. These containers host both the website and the protocol emulation. The system operates on an Ubuntu 22.04 desktop setup with Honeyd and Docker, and the LLM runs on an RTX-3090 Ti. Network traffic is managed by a router that directs specific traffic to the host machine or the virtual hosts managed by Honeyd. Traffic on ports 502, 80, and 443 are routed to the host machine, where HAProxy~\cite{haproxy} then redirects web traffic to ensure secure communication, directing HTTP to HTTPS and handling port-specific traffic.

The replication of the PLC website involves manually downloading the web server files. A Python Flask application mimics the site's functionality, including a REST API, to provide dynamic web content \cite{flask}. We manually generate the web server's SSL certificate using the OpenSSL tool~\cite{openssl}. For data logging and analysis, we employ MongoDB~\cite{mongodb} with the Beanie ODM~\cite{beanie} for robust data management. The system logs detailed information about each interaction, such as IP addresses, web pages visited, and the timing of requests and responses. It also captures login credentials attempted during brute force attacks and records incoming and outgoing Modbus server traffic.

\subsection{Interaction Analysis}
\begin{figure}[!t]
  \includegraphics[width=\linewidth]{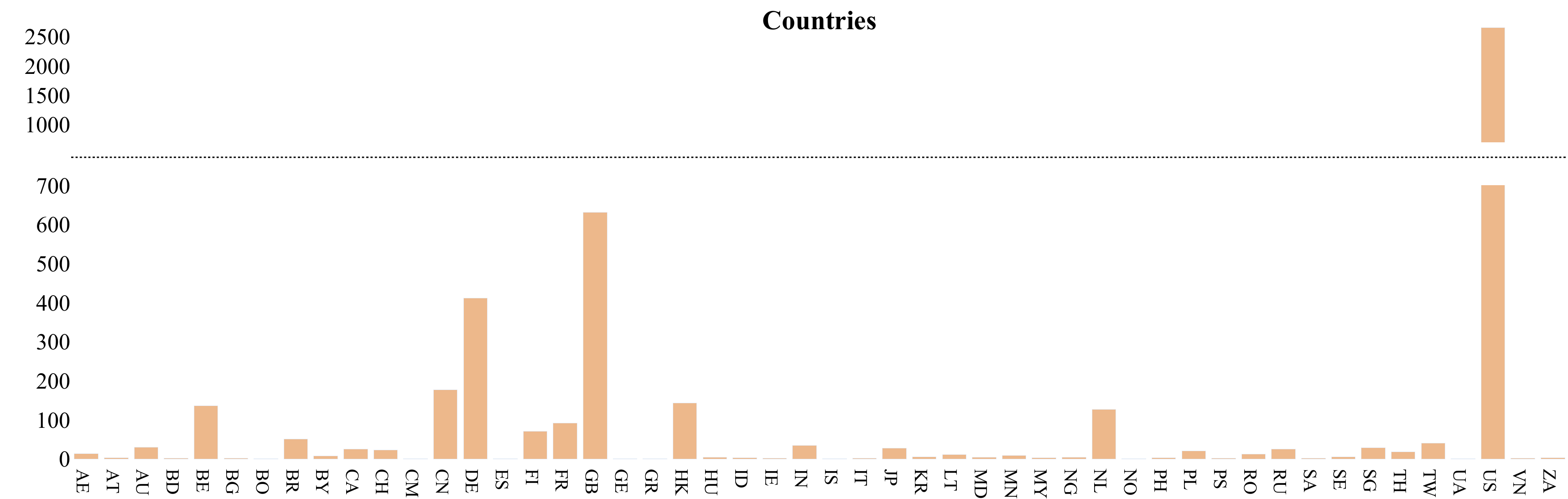}
  \caption{Geolocation of Honeypot Interaction}
  \label{fig:interaction}
\end{figure}

We exposed the developed \shortacronym~honeypot to the internet to monitor network activities. Over two months, the honeypot recorded 4895 connection attempts from unique IP addresses. Figure~\ref{fig:interaction} indicates that most of these attempts were from the United States (1399 interactions). Great Britain and Germany are the second most common sources with 632 and 412 attempts, respectively. The interaction analysis further revealed a total of 7010 requests sent to \shortacronym, where 1891 of these were valid Modbus requests. Among the valid requests, 1752 were identified to use specific FCs: 1, 3, 5, 6, 15, 16, 17, 43. To be specific, 1629 requests focused only on acquiring the device information (FC 43). Overall, \shortacronym~managed to produce 1496 valid Modbus responses.

\begin{takeaway}\label{takeaway:honeypot}
\shortacronym's capability to handle and respond to a variety of Modbus FCs from the internet demonstrates its effectiveness in emulating a honeypot and shows that LLMs can be used for ICS threat detection and analysis.
\end{takeaway}

We further utilized the AbuseIPDB~\cite{abuseipdb} database to identify the IP addresses that have been associated with malicious activity online by using a threshold value of 50 in the given scale of abuse confidence score (0 to 100). Among the total requests targeting our honeypot, 731 distinct IPs interacted more than once using the Modbus protocol, where 441 IPs were categorized as malicious. As part of our analysis, it was also noticed that 164 code injection attempts (SQL, LDAP, etc.) were made using the ``HTTP\_X\_FORWARDED\_FOR'' http protocol header or the username/password fields in the PLC login web page.

\begin{takeaway}\label{takeaway:honeypot-interactivity}
The high interactivity of attackers further proves the realism of the developed~\shortacronym~in emulating a real ICS device.
\end{takeaway}

\subsection{Resilience to Reconnaissance Tools}
To evaluate the effectiveness of the honeypot in emulating real network devices, it was tested using various network reconnaissance tools. Nmap testing revealed that the honeypot effectively camouflages as a legitimate device; for instance, Nmap shows confidence of 95\% for WAGO's PLC operating system, while \shortacronym~obtained an 87\% confidence rating. Further testing was done using Shodan~\cite{shodan}, which evaluates devices based on characteristics such as the number of open ports and specific network configurations to determine if a device is potentially a honeypot. Our deployed honeypot was successfully disguised and not labeled as a honeypot by Shodan.

\subsection{Real-time Responsiveness}
The LLM response needs to be generated in a time comparable to the real PLC response time. \shortacronym~produces a valid response in 160ms using an RTX3090Ti GPU. However, we observed lower response times, around 110ms to 140ms, when serving multiple requests in parallel using an A100 GPU. This is comparable to the reported times of real PLCs  in Honeyvp~\cite{you2021honeyvp}, which are approximately 100ms. The small difference between the two is expected to be masked by the much longer delays of the full stack network communication over public internet.~\shortacronym~achieved a maximum performance of 30 requests/sec using one Nvidia A100 GPU with a multi-instance setup when simulating a variety of environments from the literature (ICSPatch~\cite{rajput2023icspatch}), such as a desalination process, chemical process, airplane controller etc. that simulate complex real-world scenarios. \shortacronym’s response time was consistent in all processes, implying that performance does not directly depend on the complexity of the process, but rather on how \shortacronym~is structured (number of parameters) and the size of the reply defined by the protocol. The performance of \shortacronym~scales linearly with the number of GPUs: X Nvidia A100 GPUs can serve 30*X requests/second.

%% file: sections/10-discussion.tex
\subsubsection*{ByT5 Architecture and Resources:}
\shortacronym~uses the ByT5-small model from ByT5 series, which includes various sizes to balance computational complexity and capacity to handle complex patterns. The sizes range from 300M parameters for ByT5-small to 12.9B parameters for ByT5-XXL. Although ByT5-base with 528M performed similarly to ByT5-small in our tests, we opted for ByT5-small due to its lower computational demands.

\noindent\textbf{Fine-tuning an LLM to Emulate Multiple Protocols}
\shortacronym~uses separate ByT5 instances to emulate Modbus and S7comm protocols. Our initial experiments with one model emulating multiple protocols have shown promising results and this will be part of our future research.

\noindent\textbf{Emulating Unbounded Functions}
Another challenge encountered in exploring mathematical functions is related to exponential functions, which typically grow indefinitely. This characteristic poses significant difficulties in predicting accurate results, especially as function values become excessively large. However, in physical processes, variables are naturally constrained and do not extend infinitely.

\noindent\textbf{Protocol Configuration File}
\shortacronym~depends on a configuration file that includes data fields specific to various industrial protocols. This can be potentially automated in future work. However, this is a one-time effort per industrial protocol, and the number of industrial protocols is not very large.

\noindent\textbf{ICS Protocol Coverage} It is essential to highlight that \shortacronym~currently generates precise response packets for requests that involve functions within its training scope (FCs 1,2,3,4,5,6,15,16,17,43). However, the results may not be as precise for functions outside its training scope. Future work will incorporate automated function / attribute discovery for any protocol, so the configuration file describing all possible functions, features, attributes, and constraints can eventually be generated automatically.

\noindent\textbf{Real Time Performance Improvement} To minimize the response overhead of~\shortacronym~, the first approach could be to use a more modern model like RWKV~\cite{peng2023rwkv} which is also not reliant to a tokenizer, while other possible options which use a tokenizer could be Gemma-2~\cite{team2024gemma} and Phi-2~\cite{javaheripi2023phi} that promise faster inference. Secondly, pruning can be applied to the trained model to reduce complexity. Finally, a lightweight version of a generative model can be implemented. These approaches could effectively reduce the number of parameters used, thereby minimizing inference time.

%% file: sections/11-conclusion.tex
The \shortacronym~framework is, to the best of our knowledge, the first approach to use LLMs to emulate ICS network protocols and physical processes. It provides a novel platform that researchers can use to mimic network protocols, mathematical equations, and physical processes. This platform can also be used to create a real honeypot. \shortacronym~demonstrates the LLMs potential and capabilities in industrial protocol emulation, real-time interaction, and physical process capturing. LLMs can understand and emulate industrial protocols, such as Modbus, S7comm, etc. They also exhibit potential in producing credible responses during efforts to emulate some physical process. Future work will explore further automating the ICS protocol configuration file development, as well as further pushing the boundaries of the fidelity of the physical value responses when emulating a process.

%% file: sections/appendix.tex
\subsection{Additional results on protocol emulation}\label{appendix:protocol_emulation}
\begin{figure}[!t]
\vspace{-0.8em}
\subfloat[train/validation loss on Modbus]{\includegraphics[width=0.5\linewidth]{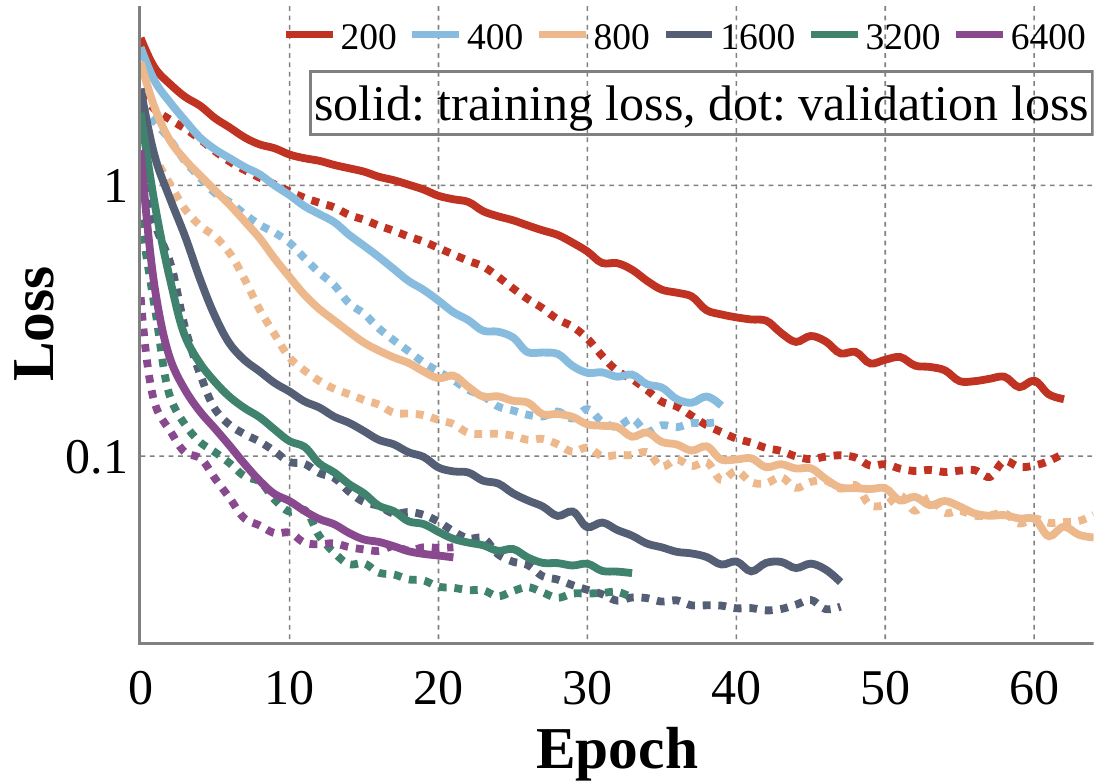}\label{subfig:appendix:protocol_emulation:mbtcp:loss}}
\subfloat[train/validation loss on s7comm]{\includegraphics[width=0.5\linewidth]{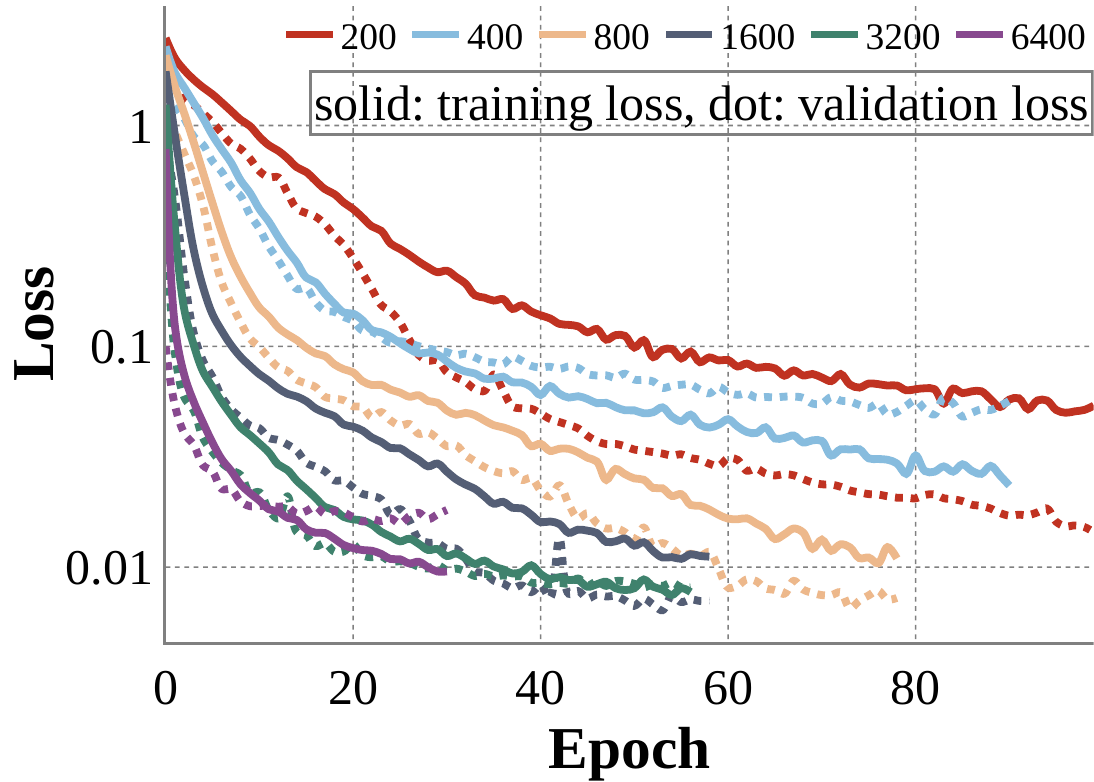}\label{subfig:appendix:protocol_emulation:s7comm:loss}}
\caption{The~\bca~and~\rva~per epoch of the byt5-small model when using different dataset sizes and protocols to finetune. A patience value of 10 epochs was used to stop the finetuning in case the validation loss was not improving, thus the different ending epochs for each dataset size.}
\label{fig:appendix:protocol_emulation:performance}
\end{figure}

In Figure~\ref{subfig:appendix:protocol_emulation:mbtcp:loss} and Figure~\ref{subfig:appendix:protocol_emulation:s7comm:loss}, show the training and the validation loss plots. For smaller datasets, overfitting is evident as depicted by the validation loss surpassing the training loss. Additionally, losses for these smaller datasets remain relatively high, indicating inadequate fine-tuning.

\subsection{Additional results on industrial testbed emulation}\label{appendix:testbed}
\begin{figure}[!t]
\vspace{-0.8em}
\subfloat[dataset size: 1600]{\includegraphics[width=0.5\linewidth]{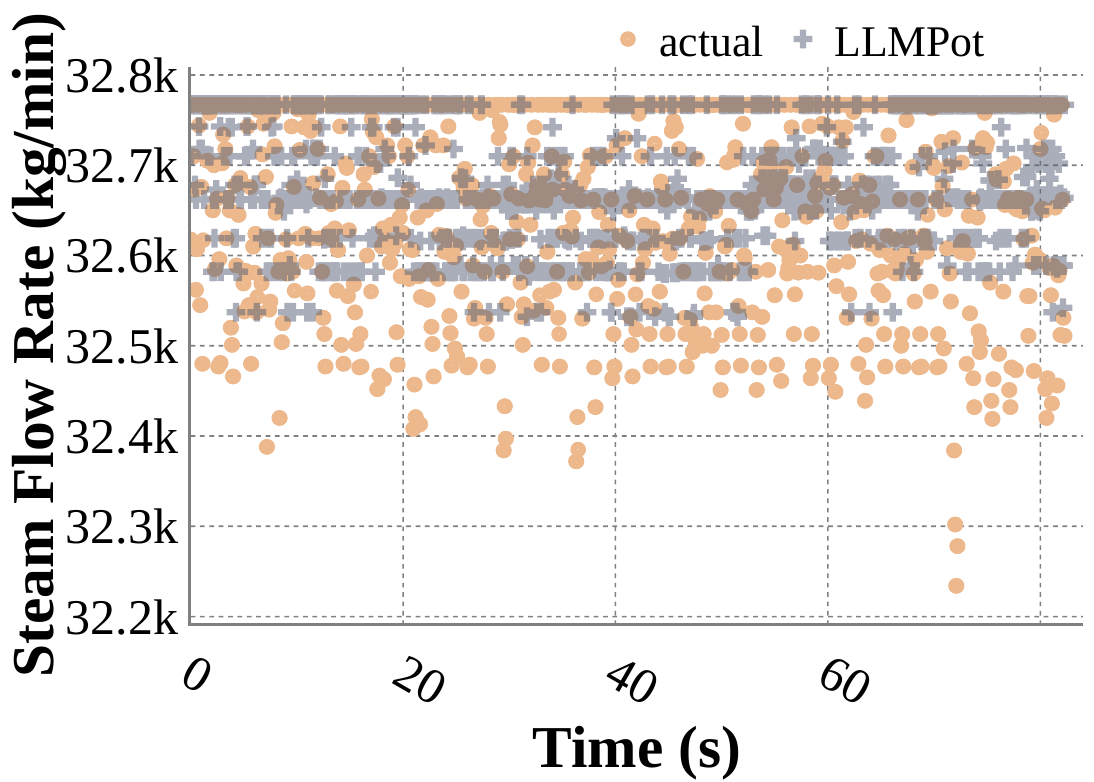}\label{subfig:testbed:sp65:s1600}}
\hfill
\subfloat[dataset size: 3200]{\includegraphics[width=0.5\linewidth]{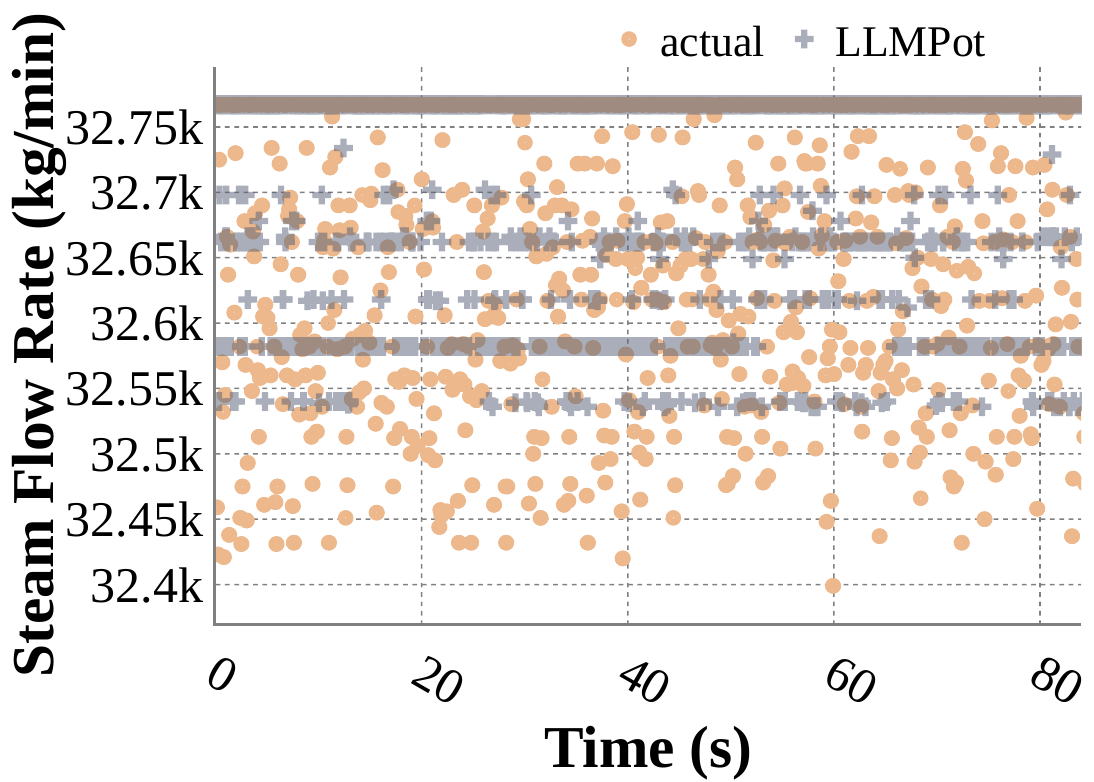}\label{subfig:testbed:sp65:s3200}}
\caption{Fluctuation of steam flow rate over time using set-point value 65 and dataset sizes, 1600 and 3200.}
\label{fig:testbed:time-sp65}
\end{figure}

Figure~\ref{fig:testbed:time-sp65} depicts the point where LLMPot fails to capture the disturbance fluctuation due to the vast amount of fluctuating points and instead averages the data points. This case is observed in all set-point values below 70 and both dataset sizes, 1600 and 3200.

Through experimentation, it was noted that with abnormally low set-point values, below 70, \shortacronym~starts to fail. It was observed that the high number of fluctuating points affected by low set-point values make it challenging for \shortacronym~to capture the disturbance fluctuations accurately and thus resulted in the model averaging the data points and losing the Ws steady-state value as shown in Figure~\ref{fig:testbed:time-sp65}. This figure depicts the point where LLMPot fails to capture the disturbance fluctuation due to the vast amount of fluctuating points. Since the ultimate objective is to emulate the physical process and to accurately reflect the values, a trade-off was made to make LLMPot accurately emulate and maintain the Ws steady-state value. 

\begin{figure}[!t]
\vspace{-1.2em}
\subfloat[setpoint: 75, dataset size: 1600]{\includegraphics[width=0.5\linewidth]{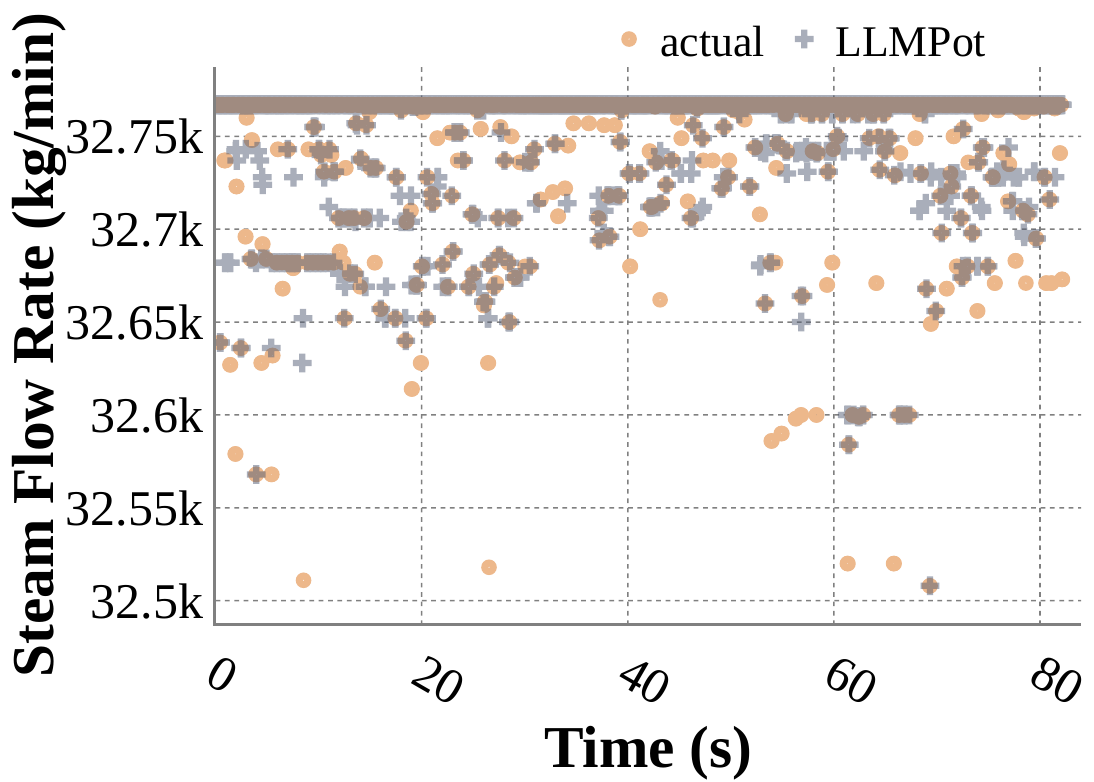}\label{subfig:testbed:sp75:s1600}}
\subfloat[setpoint: 75, dataset size: 3200]{\includegraphics[width=0.5\linewidth]{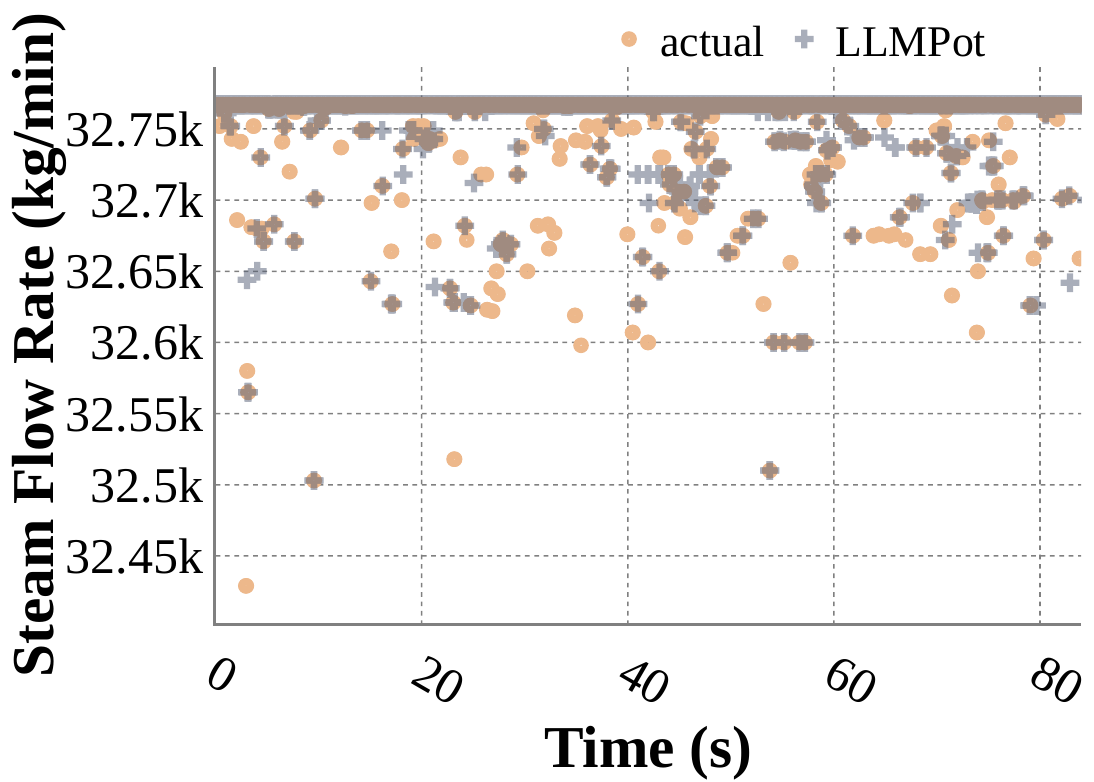}\label{subfig:testbed:sp75:s3200}}
\hfill
\subfloat[setpoint: 80, dataset size: 1600]{\includegraphics[width=0.5\linewidth]{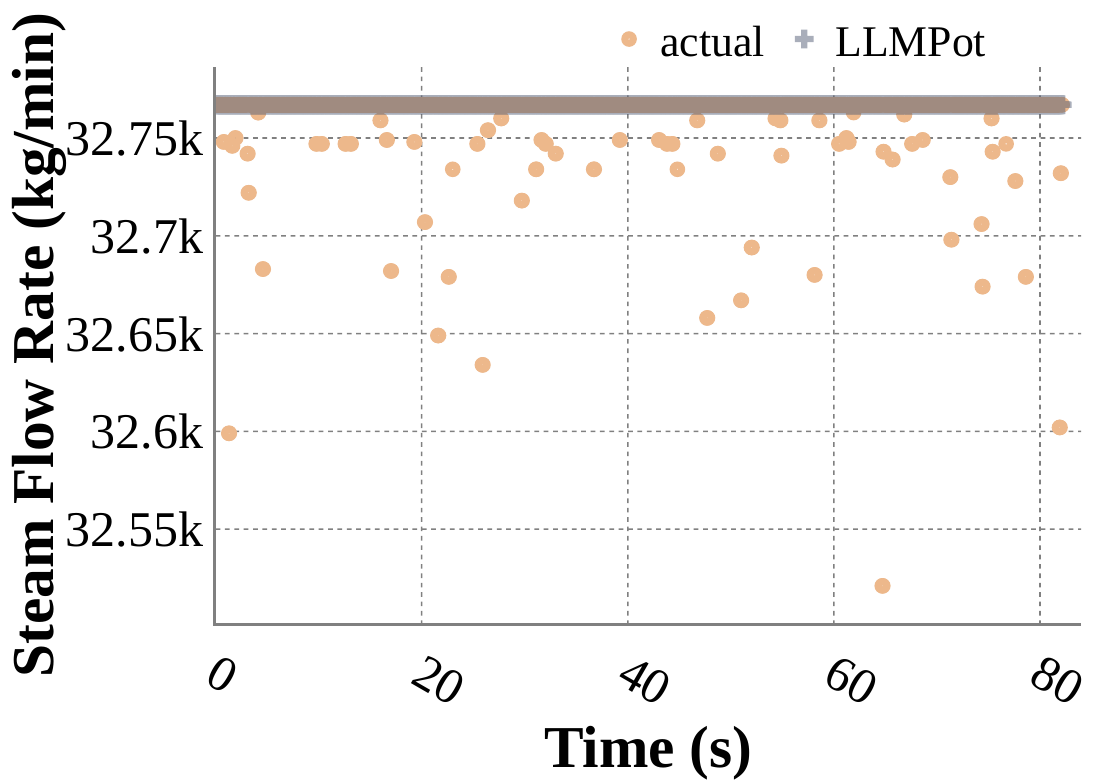}\label{subfig:testbed:sp80:s1600}}
\subfloat[setpoint: 80, dataset size: 3200]{\includegraphics[width=0.5\linewidth]{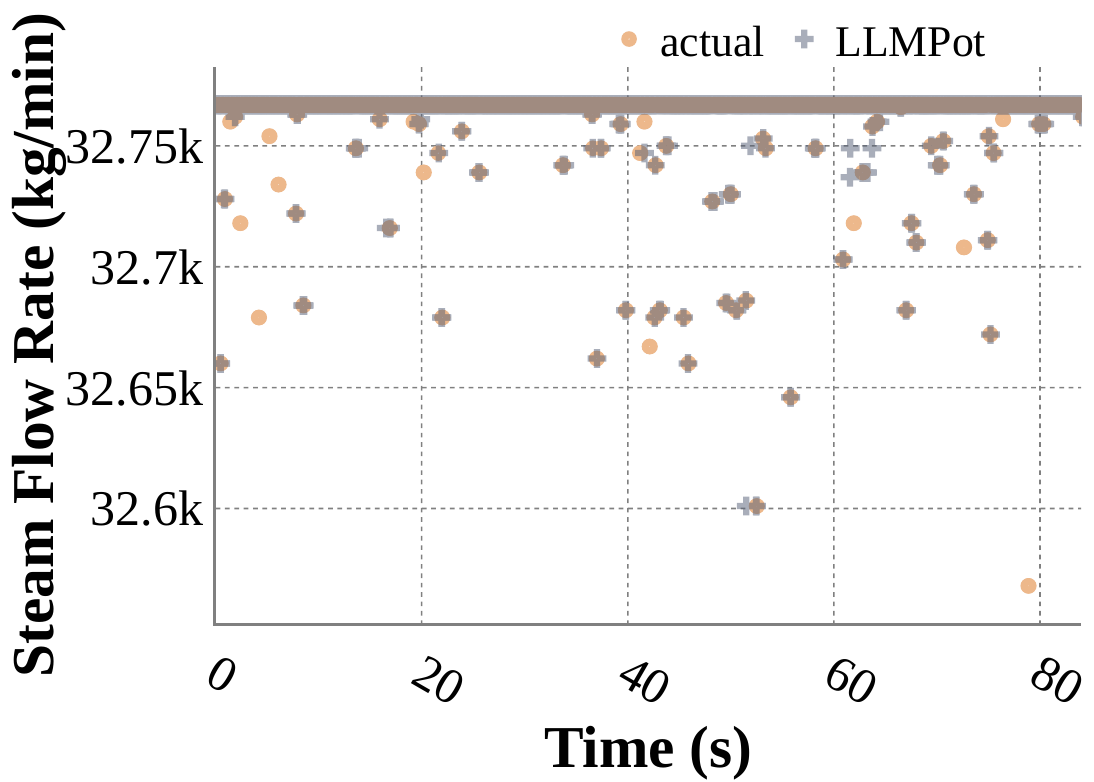}\label{subfig:testbed:sp80:s3200}}
\hfill
\subfloat[setpoint: 85, dataset size: 1600]{\includegraphics[width=0.5\linewidth]{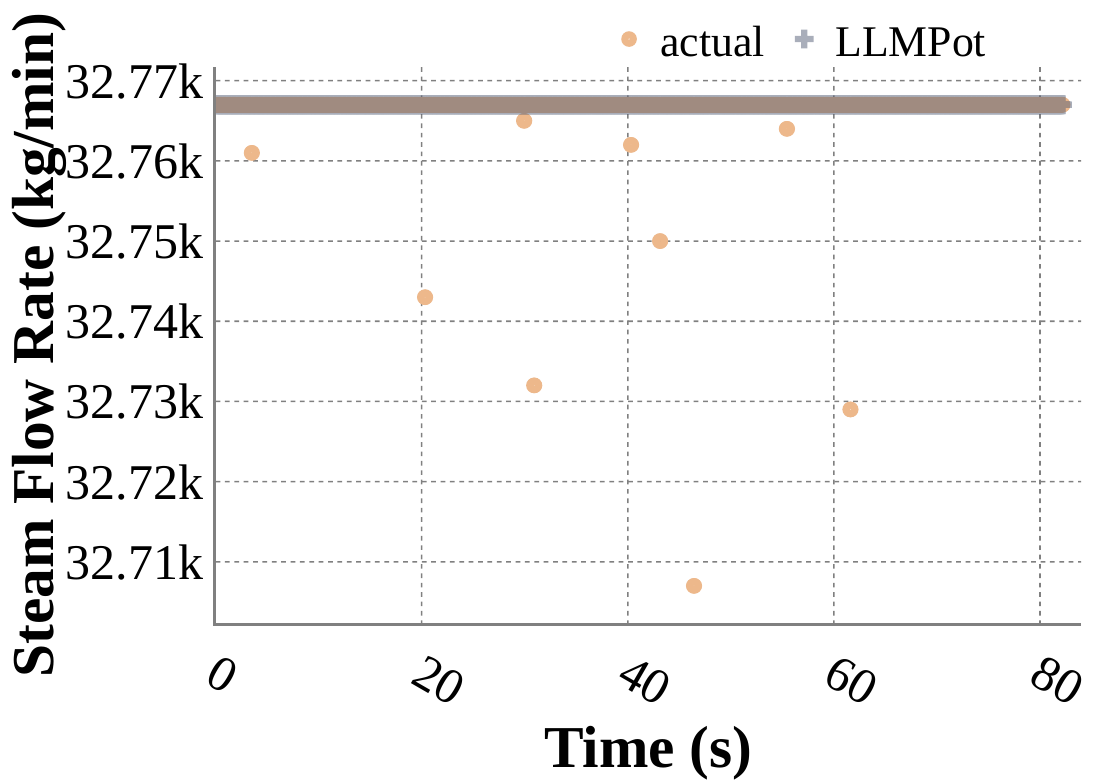}\label{subfig:appendix:testbed:sp85:s1600}}
\subfloat[setpoint: 85, dataset size: 3200]{\includegraphics[width=0.5\linewidth]{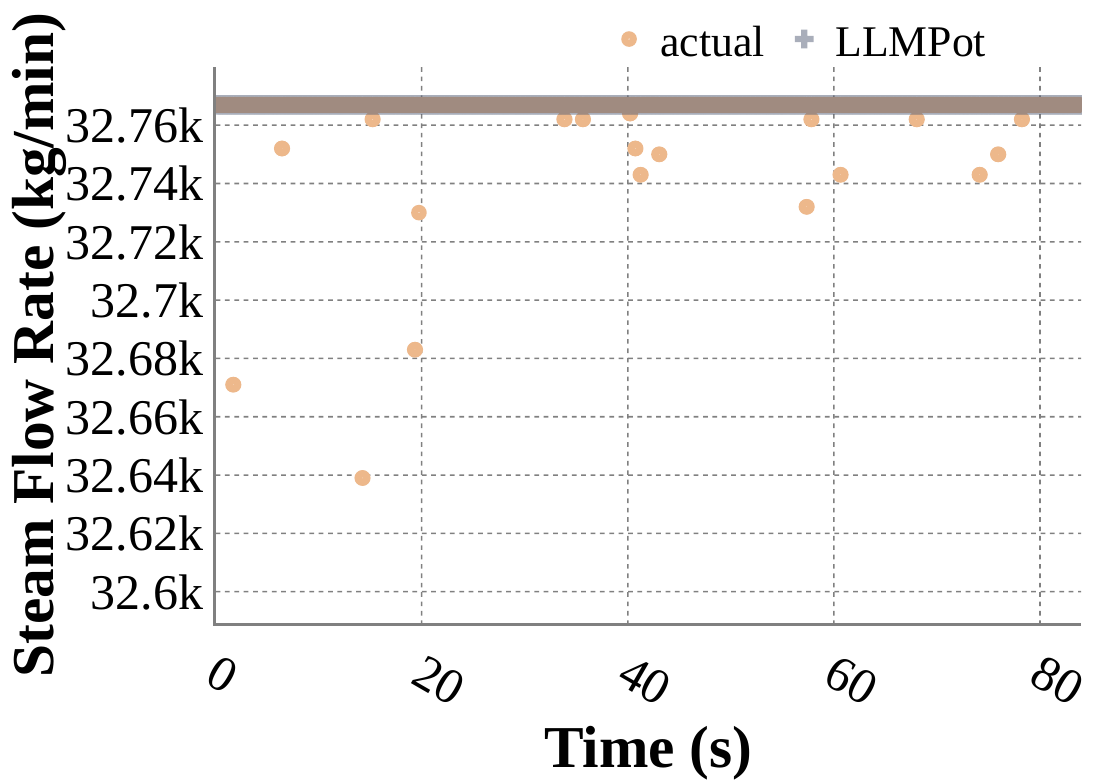}\label{subfig:appendix:testbed:sp85:s3200}}
\caption{Fluctuation of Steam Flow Rate over Time using different PID controller setpoints.}
\label{fig:appendix:testbed:time}
\end{figure}

Figure~\ref{fig:appendix:testbed:time} depicts the additional experiment made to capture the distribution of the data points collected from the real testbed and LLMPot's emulation results. For this experiment, different dataset sizes were chosen, 1600 and 3200. It was observed that in dataset sizes below 1600, LLMPot will start to fail as there are not enough data points in the dataset for the model to learn from. On the other hand, with dataset size beyond 3200, the model was facing challenges in capturing the vast amount of fluctuating points. Thus, both 1600 and 3200 sizes were chosen as they provide the best performance of LLMPot. Moreover, various set-point values were chosen in the experiment, 75, 80, and 85, in order to better understand the behavior of the process.

\begin{table}[!t]
\caption{Statistical Analysis for Real and Predicted Distribution using Kolmogorov-Smirnov test and presenting the P-value and KS-statistic.} \label{table:statistics}
\begin{tabular}{|c|c|cc|cc|}
\hline
\makecell{\\Run} & \makecell{\\Metric} & \makecell{sp: 75 \\ size: 1600} & \makecell{sp: 75 \\ size: 3200} & \makecell{sp: 80 \\ size: 1600} & \makecell{sp: 80 \\ size: 3200} \\
\hline
\multirow{2}{*}{1st} & P & 0.33717 & 0.05792 & 0.00008 & 0.99957 \\
                     & KS & 0.03295 & 0.03306 & 0.07938 & 0.00877 \\
\hline
\multirow{2}{*}{2nd} & P & 0.23519 & 0.09639 & 0.00008 & 1.00000 \\
                     & KS & 0.03619 & 0.03061 & 0.07938 & 0.00593 \\
\hline
\multirow{2}{*}{3rd} & P & 0.21852 & 0.00082 & 0.00008 & 0.99967 \\
                     & KS & 0.03682 & 0.05461 & 0.07938 & 0.01002 \\
\hline
\multirow{2}{*}{4th} & P & 0.37873 & 0.00000 & 0.00008 & 1.00000 \\
                     & KS & 0.03182 & 0.14043 & 0.07938 & 0.00687 \\
\hline
\multirow{2}{*}{5th} & P & 0.38359 & 0.03322 & 0.00008 & 0.99999 \\
                     & KS & 0.03170 & 0.03561 & 0.07938 & 0.00718 \\
\hline
\multirow{2}{*}{6th} & P & 0.51533 & 0.07030 & 0.00008 & 0.99999 \\
                     & KS & 0.02857 & 0.03217 & 0.07938 & 0.00721 \\
\hline
\multirow{2}{*}{7th} & P & 0.29466 & 0.05802 & 0.00008 & 1.00000 \\
                     & KS & 0.03420 & 0.03305 & 0.07938 & 0.00687 \\
\hline
\multirow{2}{*}{8th} & P & 0.62591 & 0.10248 & 0.00008 & 0.99996 \\
                     & KS & 0.02619 & 0.03029 & 0.07938 & 0.00781 \\
\hline
\multirow{2}{*}{9th} & P & 0.37877 & 0.10939 & 0.00008 & 0.99998 \\
                     & KS & 0.03182 & 0.02993 & 0.07938 & 0.00752 \\
\hline
\multirow{2}{*}{10th} & P & 0.37877 & 0.07498 & 0.00008 & 0.99990 \\
                      & KS & 0.03182 & 0.03186 & 0.07938 & 0.00815 \\
\hline
\end{tabular}
\end{table}

Table~\ref{table:statistics} depicts the statistical results obtained from the Kolomogorov-Smirnov test. The table outlines results obtained from 10 runs for different set-point values, 75 and 80, on two different dataset sizes, 1600 and 3200.